\documentclass[fleqn,usenatbib]{mnras}
\usepackage[T1]{fontenc}
\usepackage{ae,aecompl}
\usepackage{graphicx}
\usepackage{amsmath}	
\usepackage{amssymb}	
\usepackage{tablefootnote}
\usepackage{bm}
\usepackage{lscape}
\usepackage{breakcites}

\newcommand{\textsoft}[1]{\textsc{#1}}
\newcommand{\titlesoft}[1]{{\sc #1}}

\title[Shell-HVF-TARDIS]{Exploring the shell model of high-velocity features of Type Ia supernovae using \titlesoft{tardis}}
\author[B. W. Mulligan, K. Zhang, \& J. C. Wheeler]{
Brian W. Mulligan,$^{1}$\thanks{E-mail: bwmulligan@astro.as.utexas.edu}
Kaicheng Zhang,$^{1,2}$\thanks{E-mail: zkc14@mails.tsinghua.edu.cn}
J. Craig Wheeler,$^{1}$\thanks{E-mail: wheel@astro.as.utexas.edu}\\
$^{1}$Department of Astronomy, University of Texas at Austin, Austin, TX 78712, USA\\
$^{2}$Physics Department and Tsinghua Center for Astrophysics (THCA), Tsinghua University, Beijing, 100084, China
}

\DeclareRobustCommand\Msun{\unskip\ensuremath{\;\mathrm{M}_{\odot}}}
\DeclareRobustCommand\Rsun{\unskip\ensuremath{\;\mathrm{R}_{\odot}}}
\DeclareRobustCommand\Lsun{\unskip\ensuremath{\;\mathrm{L}_{\odot}}}
\DeclareRobustCommand\Ang{\unskip\mbox{\;\normalfont\AA}{}}
\DeclareRobustCommand{\kmSec}{\unskip\ensuremath{\;\mathrm{km}\;\mathrm{s}^{-1}}}

\DeclareRobustCommand{\Kelvin}{\unskip\ensuremath{\;\mathrm{K}}}

\newcommand{\BE}{\begin{equation}}
\newcommand{\EE}{\end{equation}}

\date{Accepted XXX. Received YYY; in original form ZZZ}

\pubyear{2019}

\begin{document}
\label{firstpage}
\pagerange{\pageref{firstpage}-\pageref{lastpage}}
\maketitle

\begin{abstract}
We explore the possible nature of high-velocity features in Type~Ia supernovae by presenting synthetic spectra generated from hydrodynamic models of interaction between the supernova and a compact circumstellar shell. We use \textsoft{tardis} to compute the spectra and compare model spectra to data from SN~2011fe at 2, 5, and 9 days after the explosion. We apply abundance models to the shells that consist of either a hydrogen, helium, or carbon-oxygen substrate with overall solar abundance of metals and depletion or enhancement of calcium abundance. We vary the calcium abundances in the shell and the ejecta to study the effect on the photospheric and high-velocity components of the calcium near-infrared triplet. The substrate leaves no imprint upon the spectra, but helium substrates are discouraged by demanding excessive calcium abundances. We find that we can approximately reproduce the blended high-velocity and photospheric velocity features at 2 and 5 days. The 9 day spectrum shows a distinct high-velocity component of the line. We are unable to reproduce this specific feature with the current models. We also explore helium-shell double detonation models, finding they tend to give no high-velocity features or excessively strong features. A very carefully chosen helium-shell mass might give a reasonable reproduction of the observed calcium features. We discuss the implications of our models for the nature of high-velocity features and their significance to the evolution and explosion of SN~Ia.

\end{abstract}


\defcitealias{2017MNRAS.467..778M}{MW1}
\defcitealias{2018MNRAS.476.1299M}{MW2}

\section{Introduction}\label{MZW1:sec:MZW1_intro}

Typical Type Ia supernovae (SN~Ia) show high-velocity features (HVF) prior to maximum light that are separated from the photospheric-velocity features (PVF) by $\sim$7,000\kmSec{} \citep{1999ApJ...525..881H, 2003ApJ...591.1110W, 2004ApJ...601.1019T, 2005ApJ...623L..37M, 2006ApJ...636..400Q, 2012ApJ...752L..26P, 2013ApJ...777...40M, 2013ApJ...770...29C, 2014MNRAS.437..338C, 2014MNRAS.444.3258M, 2015MNRAS.451.1973S, 2015ApJS..220...20Z, 2016ApJ...826..211Z}. Both the HVF and the PVF slow with time, but maintain this separation. The HVF are especially prominent in the near-infrared triplet of \ion{Ca}{II} (CaNIR), but are also observed in \ion{Si}{II} and other strong lines. The HVF exhibit a polarization of 0.1 -- 1 per cent \citep{2003ApJ...591.1110W, 2008ARA&A..46..433W, 2009A&A...508..229P}, requiring that the material that gives rise to the HVF be clumped or otherwise asymmetric.

Despite the near ubiquity of the HVF and their potential to reveal important information about the progenitor evolution or combustion physics, there is no accepted physical model. Proposed models fall into three broad categories: 1) material that is part of the ``normal'' ejecta but has stronger absorption features due to non-local thermodynamic equilibrium (nLTE) effects \citep{2013MNRAS.429.2127B}, 2) material ejected at high velocity during the explosion \citep{2005ApJ...623L..37M, 2006ApJ...645..470T, 2018ApJ...863..125K}, and 3) material swept up by the ejecta after the explosion \citep{2004ApJ...607..391G}. Any cause of these features must explain the polarization and the temporal evolution of both the HVF and PVF. 

\citet{2017MNRAS.467..778M}, hereafter \citetalias{2017MNRAS.467..778M}, investigated the interaction between a model SN~Ia and a compact circumstellar shell in an attempt to characterize the temporal behavior of the HVF in the CaNIR line. They employed a 1-D hydrodynamic simulation of the collision of the ejecta with shells of various mass from 0.003 -- 0.02\Msun{} to establish the density profile of the resulting collision and generated synthetic spectra assuming constant ion state and excitation temperature. The calcium abundance of the shell was taken to be a free parameter. They explored the effect of the equation of state, the explosion model, and the width, initial density profile and mass of the shell on the appearance and temporal evolution of the CaNIR line. They compared the evolution of the pseudo-equivalent width (pEW) of the CaNIR feature from their models to observational results from \citet{2015MNRAS.451.1973S} and found that the mass of the shell must be less than $0.012\pm0.004\Msun{}$.

\citet{2018MNRAS.476.1299M}, hereafter \citetalias{2018MNRAS.476.1299M}, used the models and techniques of \citetalias{2017MNRAS.467..778M} to explore the evolution of the HVF and PVF in the CaNIR feature in the well-observed SN~2011fe. They concluded that the CaNIR feature is better explained by the supernova model interacting with a shell than the model without a shell, with a shell of mass 0.005\Msun{} resulting in a somewhat better fit than shells of other masses. The evolution of the optical depth of CaNIR suggested that the ionization state of calcium within the ejecta and shell is not constant. \citetalias{2018MNRAS.476.1299M} noted that their model line profiles were intrinsically non-Gaussian and discussed associated ambiguities in observational methods for determining the presence and the pEW of the HVF by fitting multiple Gaussian profiles to the line features. They concluded that the physical properties of the HVF or other components could be misinterpreted.

The models of \citetalias{2017MNRAS.467..778M} and \citetalias{2018MNRAS.476.1299M} were unable to constrain the abundance of calcium in the ejecta or shell nor the underlying substrate of the shell. The latter could plausibly be hydrogen, helium or a mixture of carbon and oxygen. In this work we make use of \textsoft{tardis} \citep{2014MNRAS.440..387K} to generate synthetic spectra from a set of  models of \citetalias{2017MNRAS.467..778M} to explore constraints on all these factors, qualitatively comparing the generated spectra to the observed spectra of SN~2011fe at equivalent epochs after the explosion.

In Section \ref{MZW1:sec:methods}, we describe the supernova-shell interaction models used for generating the synthetic spectra, the abundance models that are applied to the supernova ejecta and the shell, and the additional inputs required for \textsoft{tardis}. In Section \ref{MZW1:sec:results} we present the generated spectra for each model and compare those spectra to that of SN~2011fe and discuss the implications of the results. We present our conclusions in Section \ref{MZW1:sec:conclusion}. Most of the models we present concentrate on a hydrogen substrate. Results
for other substrates are given in appendices.

\section{Methods}\label{MZW1:sec:methods}
\subsection{Supernova-shell interaction models}\label{MZW1:sec:shellmodels}

We use the supernova-shell interaction models of \citetalias{2017MNRAS.467..778M} as the source of velocity and density information for the supernova ejecta and the shell. These models make use of the \citet{2005ApJ...623..337G} delayed-detonation explosion model for the supernova itself, and surround the supernova with a compact (radius $\ll$ \Rsun{}) shell. The explosion model includes density, internal energy, and composition information for the supernova, though the composition is limited to groups rather than specific elements; e.g. silicon, sulfur, and calcium are all considered part of the silicon group. The explosion and shell data are used as starting conditions in a hydrodynamic simulation using FLASH. In the simulation, the shock is given time to propagate through the shell until both the shell and ejecta are expanding adiabatically. 

In this work, we use only models \#49, \#53, and \#57. Each model has a shell with an initial outer radius of 0.3\Rsun{}, uses delayed detonation model \texttt{c} for the explosion and ejecta with the gamma-law equation of state and has a saw-tooth density profile with the highest density at the edge closest to the explosion. Model \#s 49 and 53 are used only for evaluation of the compositions of helium envelopes of SNe~Ia of 0.01 and 0.02\Msun{}, respectively, described by \citet{2014ApJ...797...46S}. We have chosen model \# 57, with a shell of mass 0.005\Msun{}, as a slightly better match to SN~2011fe, though the differences between the spectra resulting from different masses of the shell is small, and thus this model is generally representative of shells with a mass near 0.01\Msun{}. We hereafter refer to the shell models based upon the mass of the shell rather than the model number for clarity.


The hydrodynamic models provide density as a function of velocity that is then used as an input to \textsoft{tardis}. The density is sampled at 256 points spanning the range of velocity of both the ejecta and shell. The spectra will be somewhat sensitive to the density profile of the ejecta, but we believe this to be a secondary effect compared to other model uncertainties. The angle-averaged density profile we have constructed from the models of \citet{2005ApJ...623..337G} gives a nearly exponential profile in keeping with models of this type. For the ejecta, the group composition from the \citet{2005ApJ...623..337G} model is used to provide a broad framework of the structure; the details of abundances for each element within each group are described in Section \ref{MZW1:sec:compositions}. 

\subsection{Compositions}\label{MZW1:sec:compositions}

For the composition of the ejecta, we use the group composition information that is given in the initial explosion model and the result of the hydrodynamic simulation of the interaction between the supernova and shell. In order to provide details of the individual elements of each group, we use the nucleosynthetic yields by mass of the delayed-detonation SN~Ia explosion model N100 of \citet{2013MNRAS.429.1156S} for stable nuclides. We hereafter refer to this abundance model as ``Seitenzahl-like.'' We find the relative abundance of each element within each group, then assume that the ratio holds for any areas in which elements within that group appear. The ratio of abundances for each element within the associated group are listed in Table \ref{tab:groupcompositions}. In addition to a pure Seitenzahl-like composition, we also consider composition models in which the calcium content is depleted. These models named as ``N100$\pm X$,'' where $\pm X$ is the enhancement ($+$) or depletion ($-$) of calcium by $X$~dex within the silicon group. For example, a Seitenzahl-like composition that is depleted by 1~dex in calcium would be named `N100-1.' We note that the use of stable nuclides results in excess iron that should instead be in the form of cobalt and / or nickel at the epochs that we are considering. We have determined that this overabundance of iron does affect the flux by a factor of $\sim$2 blueward of about 5000\Ang{}, but does not otherwise change the results and conclusions presented here.

\begin{table}
\caption[Group abudances for \citet{2013MNRAS.429.1156S} N100]{\label{tab:groupcompositions}The abundance by mass fraction for stable nuclides within each nuclide group for the \citet{2013MNRAS.429.1156S} N100 model.}
\begin{center}\begin{tabular}{c|c|c}
\hline
 &  & Mass\\
 &  & Fraction\\
 &  & within\\
Group & Element & Group\\
\hline
C & C & 1.0\\
\hline
O & O & 1.0\\
\hline
Mg & F & $2.2\times10^{-9}$\\
Mg & Ne & $1.8\times10^{-1}$\\
Mg & Na & $1.9\times10^{-3}$\\
Mg & Mg & $7.8\times10^{-1}$\\
Mg & Al & $3.4\times10^{-2}$\\
\hline
Si & Si & $0.66\times10^{-1}$\\
Si & P & $1.3\times10^{-3}$\\
Si & S & $2.6\times10^{-1}$\\
Si & Cl & $4.9\times10^{-4}$\\
Si & Ar & $4.5\times10^{-2}$\\
Si & K & $2.6\times10^{-4}$\\
Si & Ca & $3.4\times10^{-2}$\\
\hline
Fe & Sc & $2.5\times10^{-7}$\\
Fe & Ti & $4.3\times10^{-4}$\\
Fe & V & $1.6\times10^{-4}$\\
Fe & Cr & $1.2\times10^{-2}$\\
Fe & Mn & $1.1\times10^{-2}$\\
Fe & Fe & $8.9\times10^{-1}$\\
Fe & Co & $6.4\times10^{-4}$\\
Fe & Ni & $8.9\times10^{-2}$\\
Fe & Cu & $5.5\times10^{-7}$\\
Fe & Zn & $3.7\times10^{-6}$\\
Fe & Ga & $4.8\times10^{-13}$\\
\hline
\end{tabular}
\end{center}
\end{table}

The composition of the material causing the HVF (i.e. the shell) is unknown, so we explore many possible models to identify those that look most similar to SN~2011fe. We base all compositions on the solar abundance of metals given in \citet{2009ARA&A..47..481A}. Allowing that the material in the shell may be the result of mass transfer from a white dwarf or otherwise stripped-envelope star, we also consider compositions in which hydrogen has been completely converted to helium and compositions in which all hydrogen and helium have been converted to carbon and oxygen. We refer to the primary constituent of each composition as the substrate and to this group of compositions as ``solar-type.'' While the basis of each of the metal compositions is solar, we consider enhancement or depletion of calcium relative to a solar abundance. When referring to these abundance models, we name them by their substrate, basis composition, and calcium enhancement or depletion in dex. For example, a hydrogen substrate with solar abundance of metals and calcium enhanced by 2~dex is ``H-Solar+2.'' 

In addition to solar-type abundances, we also consider the abundances of helium envelopes that have undergone detonation just prior to the supernova explosion, as described by \citet{2014ApJ...797...46S}. We select envelopes with masses of 0.005, 0.01, and 0.02\Msun{} and the subsequent yield after detonating around a 1\Msun{} carbon-oxygen white dwarf. The masses of these envelopes span the estimated range of mass of the high-velocity material and are enhanced in silicon and / or calcium, offering a possible source of HVF. The compositions for these envelopes after detonation, derived from \citet[Figure 11]{2014ApJ...797...46S} are given in Table \ref{tab:abdmodelsshen}. When generating spectra with these abundance models, we use the models of \citetalias{2017MNRAS.467..778M} with a shell of appropriate mass, i.e. for an envelope of mass 0.01\Msun{}, we use the \citetalias{2017MNRAS.467..778M} model with a shell of mass 0.01\Msun{}. The shell masses given in \citet{2014ApJ...797...46S} do not contain the mass of outer unburned helium. The mass fractions we use for a given shell mass may thus be somewhat too large if normalized to the total mass of the shell. We also note that the kinematics of a sub-Chandrasekhar mass detonation and subsequent interaction with the envelope would not match that of the \citet{2005ApJ...623..337G} delayed-detonation supernova interacting with the shells of \citetalias{2017MNRAS.467..778M}. The resulting structure in our models has most of the density that will contribute to the spectrum of the shell near the base of the shell where the calcium that contributes to the spectrum will reside even if the original shell is layered. The detailed composition in the outer parts of the shell, helium or calcium, does not contribute much since they are accelerated to very high velocities and very small optical depths. We consider our treatment a first approximation to the effect of the helium envelope upon the spectra; the velocity of the material within the remnants of the envelope are likely to have a lower velocity, and perhaps higher density, than the shell models of \citetalias{2017MNRAS.467..778M}. 

\begin{figure}\centering
\includegraphics[height=\columnwidth,angle=-90]{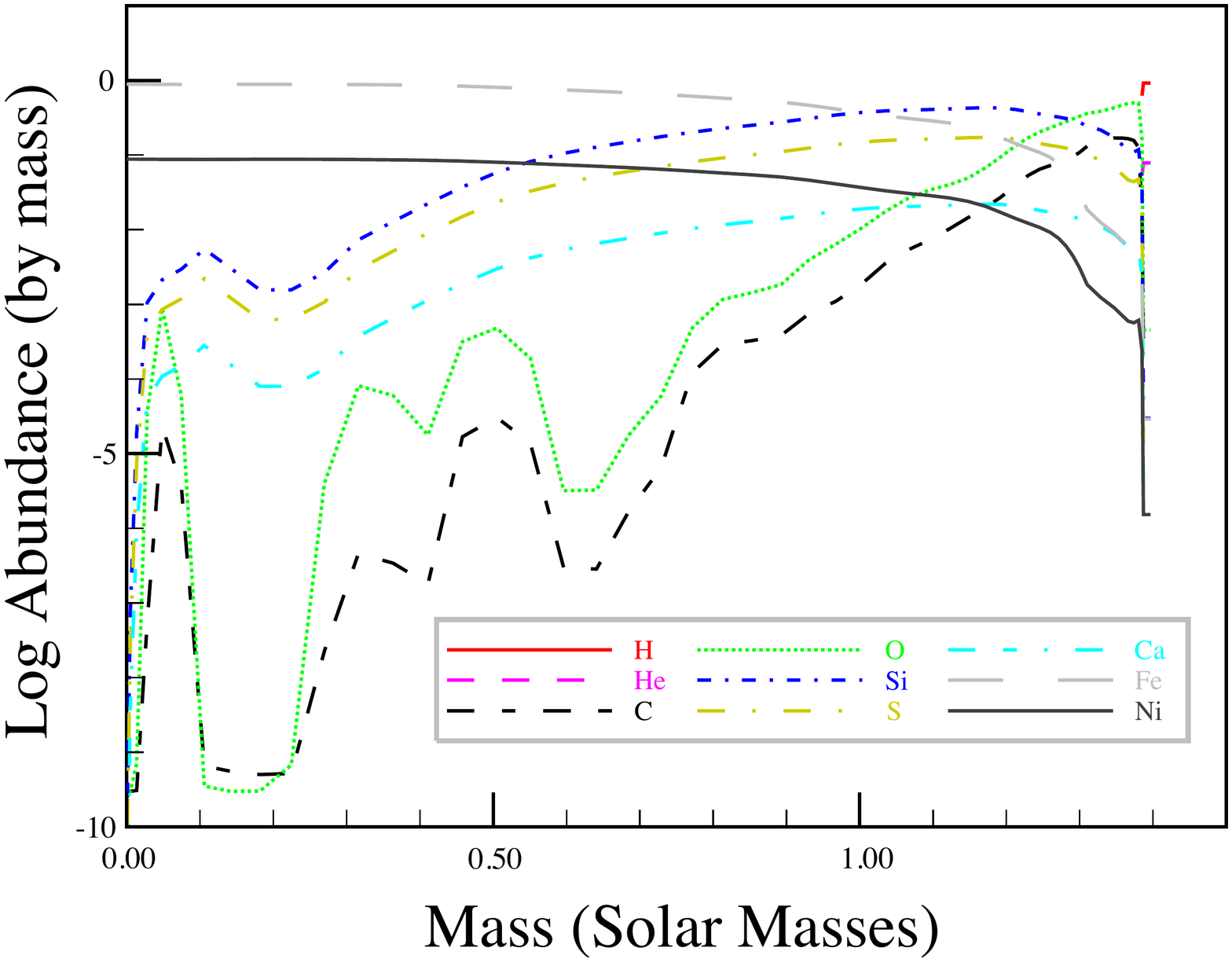}
\center
\includegraphics[height=\columnwidth,angle=-90]{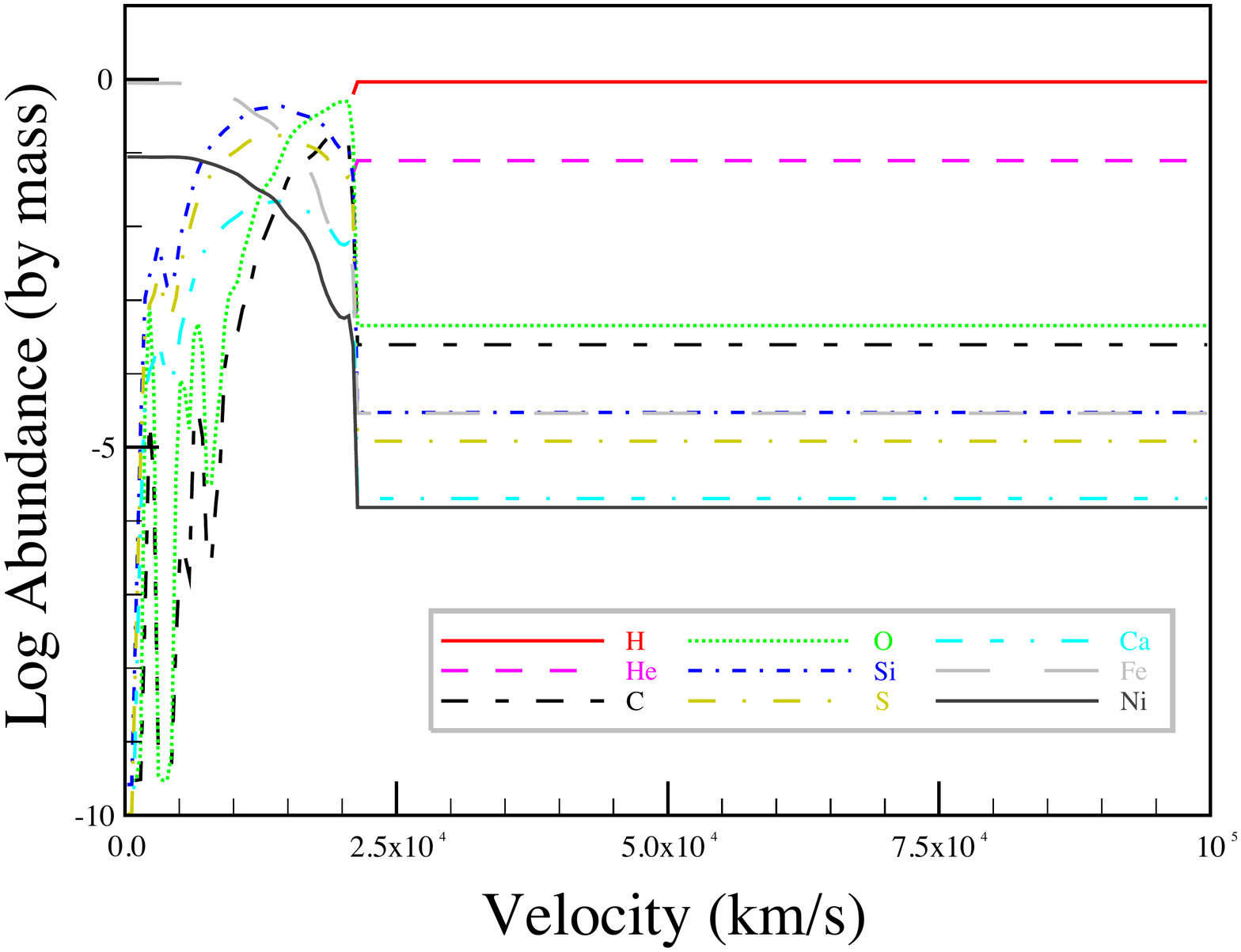}
\caption[Sample abundance by mass]{\label{fig:abund_solar_mass} Top: The abundance by mass fraction as a function of interior mass of the supernova ejecta and shell for a select set of elements, using the \citet{2005ApJ...623..337G} delayed-detonation explosion that has interacted with a shell of mass 0.005\Msun{}. This model incorporates the \citet{2013MNRAS.429.1156S} N100 model for ratios of individual elements within each nuclide group (N100+0), and a solar abundance for the material within the shell (H-Solar+0). Because we use the stable nuclides, iron is extremely enhanced within the supernova ejecta. Bottom: Similar to the top panel, but as a function of velocity. The contact discontinuity between the ejecta and shell lies at 20,880\kmSec{} A color figure is available online.}
\end{figure}

Figure \ref{fig:abund_solar_mass} 
shows an example abundance as a function of mass (top panel) and velocity (bottom panel) for an abundance model N100+0 for the ejecta and model H-Solar+0 for the shell. This figure demonstrates that, while the shell consists only of a small fraction of the mass of the supernova ejecta, it occupies a large range of velocities. For the shell with a mass of 0.005\Msun{}, the contact discontinuity lies at 20,880\kmSec{}.

\begin{table*}
\caption[Abundances of \citet{2014ApJ...797...46S} envelopes]{\label{tab:abdmodelsshen}Abundances by mass fraction for helium envelopes of mass 0.005, 0.01, and 0.02\Msun{} from \citet{2014ApJ...797...46S}.}
\begin{center}\begin{tabular}{c|c|c|c|c|c|c|c}
 & Envelope &  &  &  &  &  & \\
 & Mass & $\log$ & $\log$ & $\log$ & $\log$ & $\log$ & $\log$\\
Name & [\Msun{}] & $X_{\mathrm{He}}$ & $X_{\mathrm{Si}}$ & $X_{\mathrm{Ca}}$ & $X_{\mathrm{Ti}}$ & $X_{\mathrm{Cr}}$ & $X_{\mathrm{Fe}}$\\
\hline
S\&M-M0.005 & 0.005	&	-0.11 & -0.72 & -$\infty$ & -$\infty$ & -$\infty$ & -$\infty$\\
S\&M-M0.01 & 0.01	&	-0.16 & -2.61 & -1.62   & -4.16   & -$\infty$ & -$\infty$\\
S\&M-M0.02 & 0.02	&	-0.24 & -3.59 & -0.82   & -0.65   & -1.74   &  -3.59\\
\hline
\end{tabular}
\end{center}
\end{table*}


\subsection{Synthetic spectra}\label{MZW1:sec:tardis-param}

We use \textsoft{tardis} to generate the synthetic spectra of the models described in Sec. \ref{MZW1:sec:shellmodels} and using the composition described in Sec. \ref{MZW1:sec:compositions}. \textsoft{Tardis} is well suited for the particular problem explored in this paper that requires the computation of a large number of models to explore CaNIR HVF parameter space. \textsoft{Tardis} can only compute supernova atmospheres in spherical symmetry, and while the HVF problem surely requires departure from spherical symmetry, our current models are also restricted to that regime. Given that condition, \textsoft{tardis} uses Monte Carlo (MC) methods to iterate to a self-consistent calculation of the radiation field and corresponding ionization and excitation and to compute a synthetic spectrum. The radiative transfer and associated physics can be treated with various levels of sophistication. While quite general, \textsoft{tardis} was originally designed to be efficiently applied to SN~Ia, as we do here. In their presentation of the \textsoft{tardis} code, \citet{2014MNRAS.440..387K} note that of the prominent species present in the atmospheres of SN~Ia, silicon, sulfur, magnesium, and calcium, \ion{Ca}{II} features are the least sensitive to the choice of excitation mode and are well represented by simple Boltzmann excitation levels. This also bodes well for the current study.

\textsoft{Tardis} does not simulate spectral evolution directly, but provides a ``snapshot'' at a given epoch, again a process well suited for the current problem where we study conditions at three specific epochs in the expansion of of the supernova. \textsoft{Tardis} accepts an arbitrary density profile that we provide with our shell interaction models and an arbitrary abundance profile that we vary in this study. The total luminosity is specified for the supernova; \textsoft{tardis} iterates upon a photospheric luminosity that results in the desired total luminosity. We use the luminosity of SN~2011fe as reported by \citet{2013A&A...554A..27P}, using linear interpolation of the log of the luminosity at each epoch, resulting in luminosities of $\log (L/\Lsun{}) = $ 7.81 at 2~d, 8.62 at 5~d, and 9.19 at 9~d after the explosion.

\textsoft{Tardis} does not treat non-radiative energy sources such as radioactive decay and we neglect all such effects in this work. This means the photosphere must be external to the regions in which the majority of the luminosity is produced, a good approximation for the early phases we study here. The radiation field is injected at the inner boundary (the photosphere) with a blackbody temperature consistent with the luminosity adopted at that radius. For the photospheric velocity, we use the photosphere velocities determined in \citetalias{2018MNRAS.476.1299M} for the model with a shell of mass 0.005\Msun{}, smoothed by a third-degree polynomial to reduce noise, resulting in a photospheric velocity of 16,470\kmSec{} at 2~d after the explosion, 14,300\kmSec{} at 5~d after the explosion, and 11,780\kmSec{} at 9~d after the explosion. Finally, for the initial temperature estimates, we use the color indices of \citet{2016ApJ...820...67Z} to estimate the color temperature of the radiation. We select initial temperatures of 6325~K at 2~d after the explosion, 6676~K at 5~d after the explosion, and 7321~K at 9~d after the explosion. We acknowledge that the effective temperatures of SN~Ia are not well identified by the color temperature, but in practice we find that the choices of the initial radiation temperatures have little effect on the final temperatures in \textsoft{tardis}. The final radiation temperatures generated by \textsoft{tardis} at the inner boundary are 11,100 -- 11,800\Kelvin{} at 2~d, 11,600 -- 12,100\Kelvin{} at 5~d, and 13,400 -- 13,800\Kelvin{} at 9~d, with the ranges due to the slight differences in the electron density and total opacity resulting from the intermediate mass elements.

We use \textsoft{tardis} in the nebular mode for ionization and the dilute-LTE mode for excitation. Radiative rates are treated in the detailed mode and line interactions are treated in the macroatom mode. We use $10^5$ packets for models during convergence, and $5\times10^5$ or $10^6$ packets for each final spectrum. We have computed about 150 synthetic spectra using \textsoft{tardis} in the course of this study. We find that each model requires several CPU-hours of computing time rather than the few minutes per spectrum described in \citet{2014MNRAS.440..387K}. This is likely due to our use of detailed density and abundance models as inputs.

\subsection{Reference spectra of SN~2011fe}

\begin{figure}\centering
\includegraphics[height=\columnwidth,angle=-90]{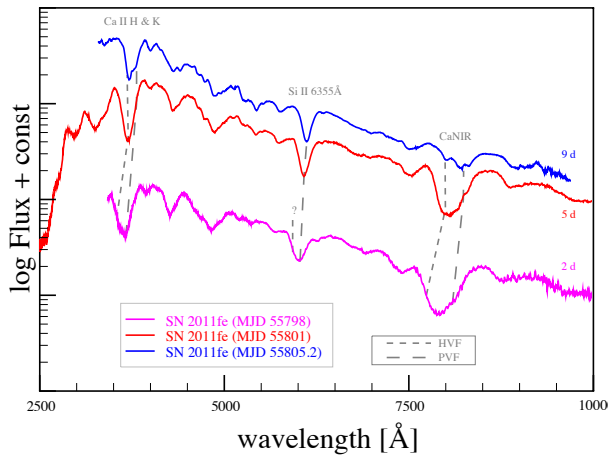}
\caption[SN2011fe examples]{\label{fig:SN2011e} Spectra of SN~2011fe at the selected epochs, highlighting the \ion{Ca}{II} H\&K, \ion{Si}{II} 6355\Ang{}, and CaNIR features. Dashed lines highlight the evolution of the PVF and HVF. The time since the explosion is listed in the right side of the figure near each spectrum. The bottom spectrum represents MJD55798 = 2d after explosion, the middle MJD55801 = 5 d and the top spectrum MJD55805.2 = 9d. A HVF in \ion{Si}{II} near 5925\Ang{} on MJD 55799, is indicated with a question mark and line at MJD 79898, but HVF have not been identified at the two later epochs that are shown here \citep{2015MNRAS.451.1973S}. A color figure is available online. }
\end{figure}

We use spectra of SN~2011fe as a comparison at each epoch (2, 5, 9~d) in order to get a sense of the relative strength of observed features, mainly focused on the \ion{Ca}{II}, \ion{Si}{II}, and \ion{O}{I} features. The spectra were acquired from \citet{2017ApJ...835...64G}, dereddened according to \citet{1989ApJ...345..245C} with corrections to the visual range from \citet{1994ApJ...422..158O} using $E(B-V)$ of 0.0077, and shifted into the rest frame wavelength using a redshift of 0.000804. We assume an explosion date of MJD~55796.696 from \citet{2011Natur.480..344N}, and use the \citet{2011Natur.480..344N} spectrum taken on MJD~55798.2 as a reference at 2~d after the explosion, the \citet{2014MNRAS.439.1959M} spectrum taken on MJD~55801.17 as a reference at 5~d after the explosion, and the \citet{2012ApJ...752L..26P} spectrum taken on MJD~55805.2 as a reference at 9~d after the explosion. At each epoch, we scale the data such that it approximately matches the synthetic spectra, typically using the blue side of the \ion{Si}{II} feature as the point of normalization. Figure \ref{fig:SN2011e} shows these spectra at each epoch, and highlights the \ion{Ca}{II} H\&K, \ion{Si}{II} 6355\Ang{}, and CaNIR HVF and PVF.

\section{Results}\label{MZW1:sec:results}

We consider each abundance model for the shell and ejecta at three epochs: 2~d, 5~d, and 9~d after the explosion, spanning the range in which the high-velocity features of calcium are significant relative to the photospheric features. At each epoch, we consider the range of calcium abundance in the shell that may have an observable affect upon the spectrum. As described in the following sub-sections, we find that the solar-type composition models with a hydrogen substrate are broadly representative of the other solar-type models, and therefore use the hydrogen substrate only to demonstrate the effect of adjusting the calcium yield within the ejecta and to generate a ``best'' fitting model. We present the results with a helium and with a carbon-oxygen substrate in Appendices \ref{MZW1:apdx:he} and \ref{MZW1:apdx:CO}, respectively.

We begin with a standard model and then alter the calcium abundance in the shell primarily to modify the blue portion of the CaNIR and the calcium abundance in the ejecta primarily to modify the red portion. For very strong shell calcium, flux is shifted to the red and also modifies that portion of the spectrum. We first illustrate variations in the shell abundance, then the ejecta abundance and then present a model that produces the overall best match to the data. This latter step is produced with ``fit by eye" rather than any formal fitting procedure, given the expense of each model atmosphere.








\subsection{2 days after the explosion}
\subsubsection{Solar-type with hydrogen substrate}

\begin{figure}\centering
\includegraphics[height=\columnwidth,angle=-90]{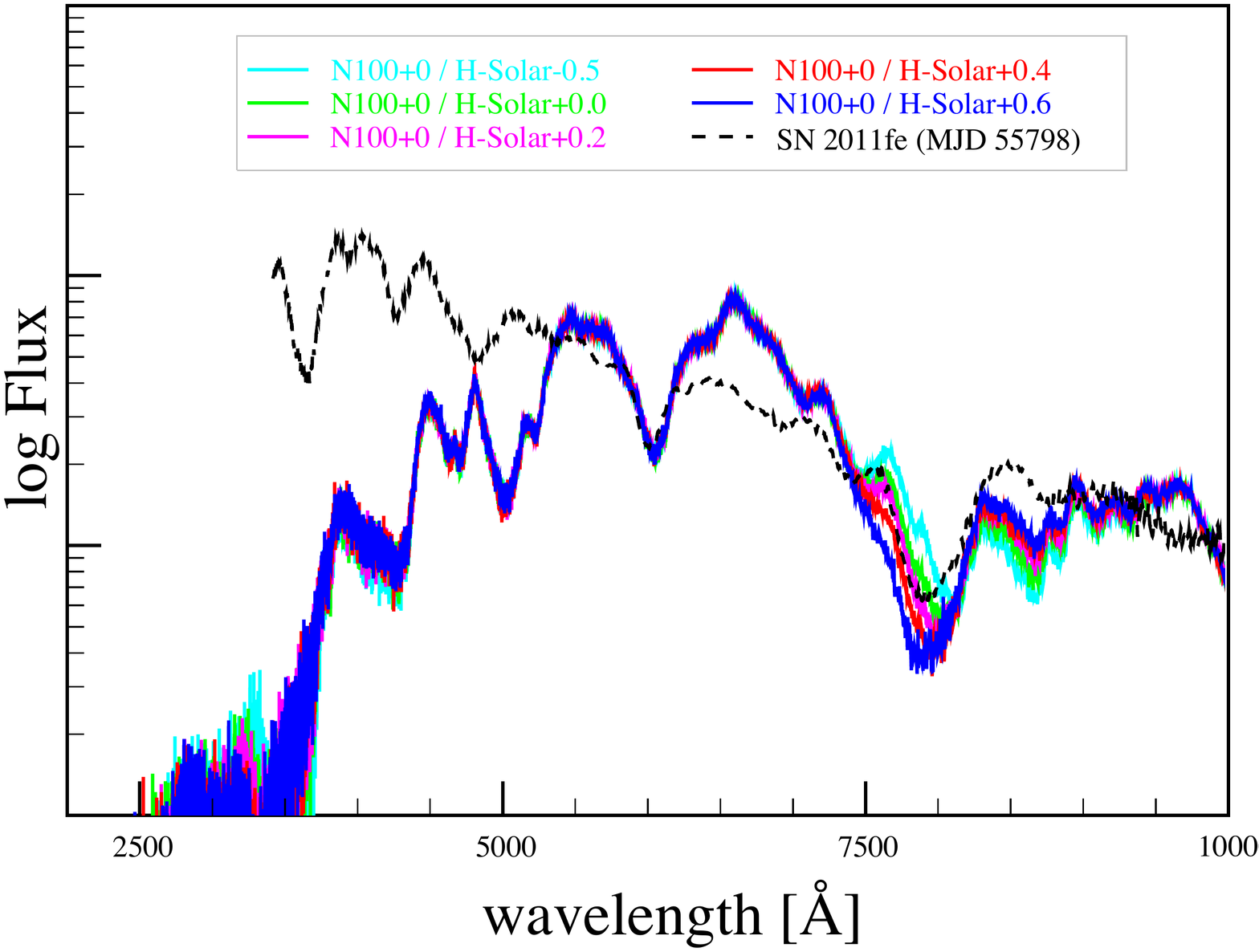}
\center
\includegraphics[height=\columnwidth,angle=-90]{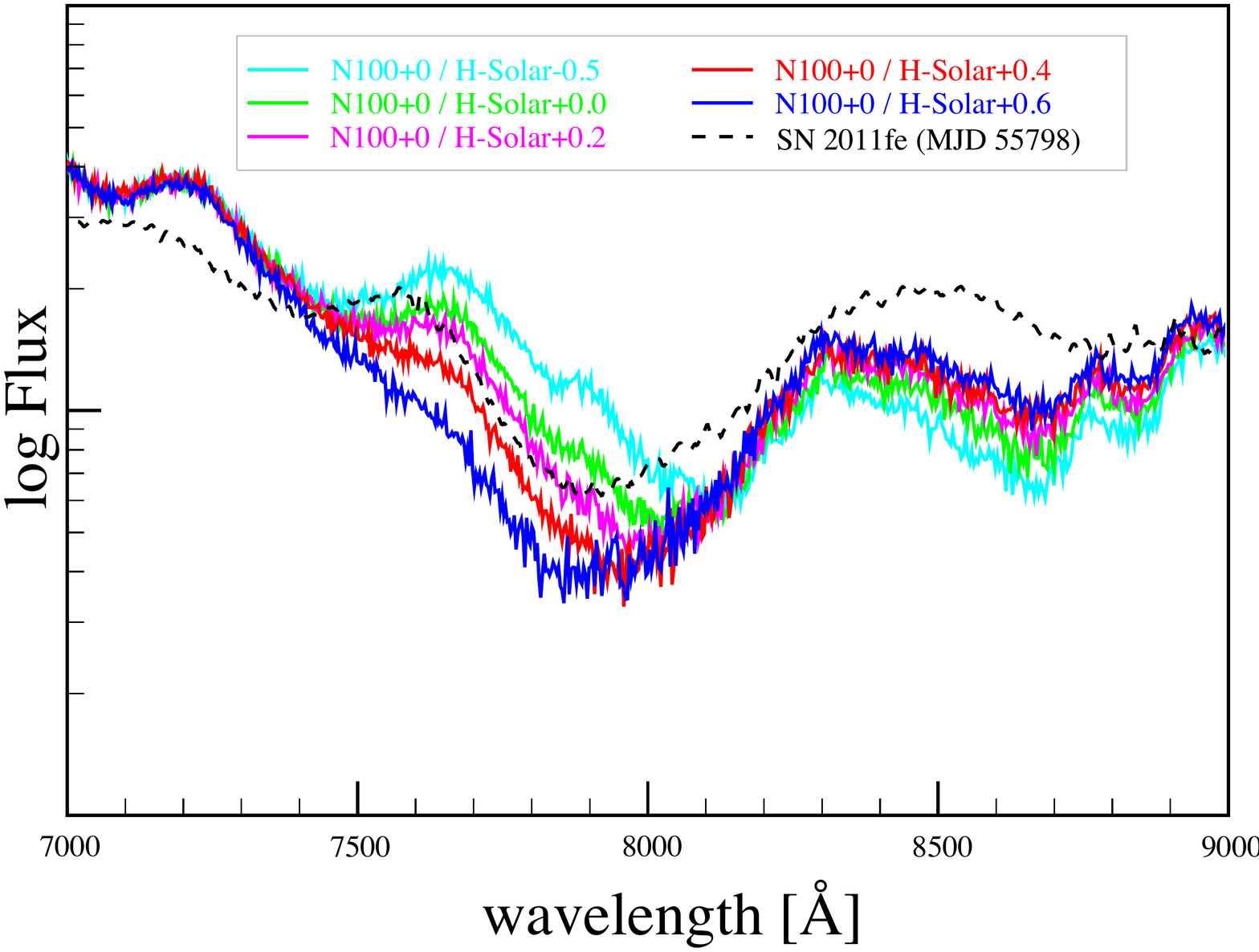}
\caption[H substrate shells at 2~d]{\label{fig:d2-H} Top: Synthetic spectra at 2~d after the explosion using a N100+0.0 abundance for the ejecta and solar-type abundances with a hydrogen substrate for the shell, with calcium depletion or enhancement ranging from -0.5 -- +0.6~dex within the shell. The spectrum of SN~2011fe at MJD~55798.2 from \citet{2011Natur.480..344N} is shown by the black dashed line for reference. Only the CaNIR feature is affected by the calcium abundance within the shell. 
Bottom: Similar to the top panel but focused on the CaNIR feature between 7000 -- 9000\Ang{}. At 7800 \AA, the models from top to bottom represent N100+0/H-solar-0.5, N100+0/H-solar+0.0, N100+0/H-solar+0.2, N100+0/H-solar+0.4, and N100+0/H-solar+0.6, respectively. This order is reversed at 8500 \AA.
A color figure is available online.}
\end{figure}

Figure \ref{fig:d2-H} shows the spectra that result from abundance model N100+0 for the ejecta and the hydrogen substrate abundance models for the shell. Note that as the quantity of calcium within the shell increases in the models, the absorption increases, leading to the P~Cygni peak also getting stronger, illustrating the redistribution of the flux. Similar behavior is seen in many of the models presented below. The \ion{Ca}{II} H\&K feature is not affected by the calcium within the shell at this epoch. 

Despite the shell in these abundance models containing mostly hydrogen, there is no evidence of hydrogen absorption or emission within the synthetic spectra, including in the mid- and far-infrared to 3$\;\mu\mathrm{m}$ (not shown in the figure). We note that limitations in the way \textsoft{Tardis} treats excitations may play a role in this context. \textsoft{Tardis} is not well designed to treat excitations in high energy levels such as the n = 2 level of hydrogen from which the Balmer lines arise. This may serve to artificially suppress the model hydrogen line strengths.

The bottom panel of Figure \ref{fig:d2-H} shows just the CaNIR feature for the hydrogenic shells. The models with -0.5~dex and with solar calcium abundance in the shell have what appears to be a discernable HVF that is not seen in the models with more calcium in the shell, nor in SN~2011fe. This feature may show up in other models (see Figures \ref{fig:d5-H}, \ref{fig:d2-He}). The model with -0.5~dex calcium abundance in the shell shows that the absorption within the ejecta (i.e. the PVF) may be too strong at this epoch when using the N100+0.0 model. The shell with only a solar abundance of calcium cannot explain the observed HVF component, but an enhancement of +0.2 -- +0.4~dex above solar value of calcium within the shell, with the N100+0.0 composition for the ejecta, does reasonably well in fitting the observed feature. As the quantity of calcium within the shell increases, the feature gets deeper and broader toward the blue, while the P Cygni peak is enhanced. An enhancement of 0.6~dex or more in calcium results in an excessively strong CaNIR feature.

There is an additional model absorption feature that appears near $\sim$8600\Ang{} that has not been observed (or at least has not been observed to be as strong) in SN~Ia and appears to get weaker with increasing calcium content in the shell. This feature will appear in nearly all spectra generated from the compositions that we consider. The apparent decrease in strength of this feature with increasing calcium within the shell is largely the result of increasing P~Cygni emission as the CaNIR feature gets stronger. Fine tuning of the structure or composition of the ejecta would assist in improving the appearance of this feature but that is not the focus of this work.

It is notable that the synthetic spectrum is very deficient in the blue and may have excessive P~Cygni emission associated with the silicon 6355\Ang{} feature. The lack of blue photons may be the result of the use of the stable nuclides from the Seitenzahl N100 model --- this particular mix only occurs years after the supernova when radioactive nuclei have decayed to stable isotopes; most notably all nickel-56 has decayed to iron-56; as the iron-group elements make up one per cent or more (by mass) of most of the ejecta, there is a strong excess of iron throughout the ejecta that causes absorption, particularly in the blue. At 2~d after the explosion, over 99 per cent of the material that will become iron-56 is still either nickel or cobalt. Not accounting for radiative effects, this suggests that the iron absorption is about 100 times too strong in these synthetic spectra. An alternative is that there may be limitations in the manner in which \textsoft{tardis} determines the photosphere. In a ``scattering" code such as \textsoft{tardis}, the choice of photospheric location can potentially affect the overall spectral energy distribution. We have explored this possibility by varying the photospheric radius by 10\%, with little effect. A larger variation would displace the minima of the photospheric features and hence violate that strong constraint. A deeper exploration of this issue may be warranted, but is beyond the scope of this paper.

\begin{figure}\centering
\includegraphics[height=\columnwidth,angle=-90]{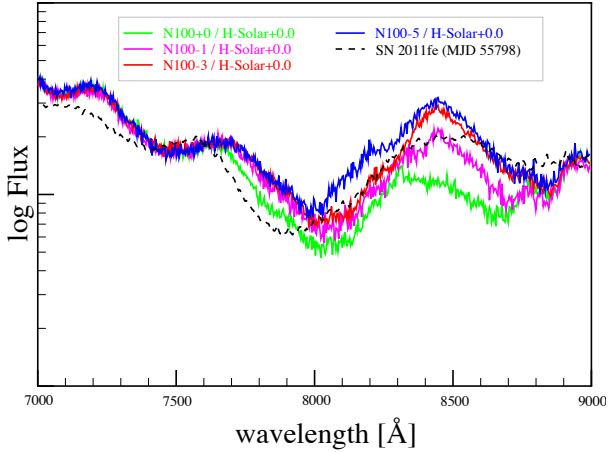}
\caption[Varying ejecta composition at 2~d (CaNIR)]{\label{fig:d2-Seit} Like the bottom panel of
Figure \ref{fig:d2-H} at 2~d after the explosion, but showing the effect of depletion of calcium within the ejecta with a solar abundance of calcium in the shell. The ejecta has a Seitenzahl-like composition, with calcium yield adjusted by -5 -- 0~dex. At 8500 \AA, the models from bottom to top represent N100+0/H-solar+0.0, N100-1/H-solar+0.0, N100-3/H-solar+0.0, and N100-5/H-solar+0.0, respectively.
Depletion of the calcium within the ejecta results in modest reduction in the strength of the CaNIR feature at this epoch. There is no clearly preferred calcium abundance within the ejecta at this epoch -- a slight depletion of the calcium results in a better fit near 8400\Ang{}, but stronger depletion results in a better fit near 8100\Ang{}. A color figure is available online. }
\end{figure}

Figure \ref{fig:d2-Seit} shows the effect of calcium within the ejecta on the synthetic spectra, depleting the calcium yield relative to the \citet{2013MNRAS.429.1156S} N100 model with stable nuclides by -5 -- 0~dex. Decreasing the calcium within the ejecta has an effect of slightly weakening the CaNIR feature, though increasing the flux of the related P~Cygni peak. The effect of the ejecta upon the CaNIR feature is entirely on the red side of the feature, at wavelengths $\gtrsim$8000\Ang{}. It is not clear that any one model of abundance in the ejecta results in a spectrum that is a better fit to the observed CaNIR feature at this epoch. Replacement of the calcium within the ejecta with silicon results in a slight enhancement of the \ion{Si}{II} 6355\Ang{} feature.

\begin{figure}\centering
\includegraphics[height=\columnwidth,angle=-90]{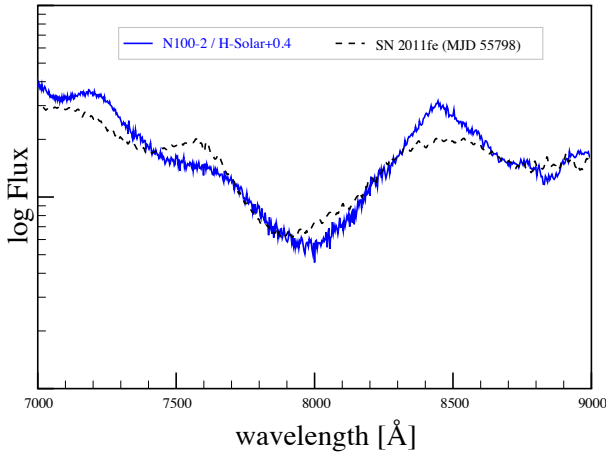}
\caption[Improved fit using H substrate at 2~d (CaNIR)]{\label{fig:d2-H-CaNIR-Best} 
Like the bottom panel of Figure \ref{fig:d2-H} at 2~d after the explosion, but for a model that represents a combination of depletion of calcium in the ejecta and enhancement of calcium in shell (-2~dex relative to the \citet{2013MNRAS.429.1156S} N100 model with stable nuclides, +0.4~dex above solar value in the shell) to better fit the observed CaNIR feature. The overall fit is reasonably good, though the P Cygni peak is slightly too strong and sharp, and the blue side of the CaNIR feature is slightly too round. A color figure is available online.}
\end{figure}

Figure \ref{fig:d2-H-CaNIR-Best} shows the result of an effort to produce a combination of ejecta and shell abundances to better match the observed CaNIR feature of SN~2011fe. We select a depletion of 2~dex of calcium relative to the \citet{2013MNRAS.429.1156S} N100 model with stable nuclides, and 0.4~dex of enhancement of calcium within the shell, relative to a solar value, based upon visual inspection of the results shown in Figures \ref{fig:d2-H} and \ref{fig:d2-Seit}. The \ion{Ca}{II} H\&K feature (not shown) is not affected by the choice of calcium within the ejecta or shell. This model captures the width and depth of the CaNIR feature reasonably well, though the peak of P~Cygni emission is slightly too sharp and has slightly too much flux. The sharpness of the P~Cygni peak may be in part due to the absorption feature near 8600\Ang{}. The blue edge of the CaNIR feature is slightly too rounded relative to the observed feature.

\subsubsection{Shen \& Moore-type envelopes}

\begin{figure}\centering
\includegraphics[height=\columnwidth,angle=-90]{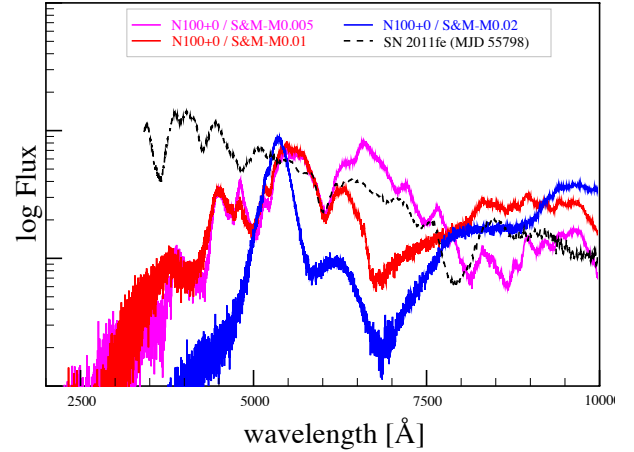}
\caption[Shen \& Moore envelopes at 2~d]{\label{fig:d2-Shen} Like Figure \ref{fig:d2-H} at 2~d after the explosion, but for a shell with \citet{2014ApJ...797...46S} type abundances. The model with a 0.005\Msun{} envelope is shown in magenta (uppermost spectrum at 7000 \AA), the 0.01\Msun{} envelope is shown in red (middle spectrum at 7000 \AA), and the 0.02\Msun{} envelope is shown in blue ( bottom spectrum at 7000 \AA). The 0.01\Msun{} model results in a deep and extended near-infrared feature; the 0.02\Msun{} model results in an extremely extended feature that blends with the \ion{Si}{II} 6355\Ang{} feature. The 0.005\Msun{} model does not contain calcium, so there is no calcium HVF; despite containing over 22 per cent silicon, there is also no evidence of a \ion{Si}{II} HVF. A color figure is available online.}
\end{figure}

Figure \ref{fig:d2-Shen} demonstrates the synthetic spectra resulting from \citet{2014ApJ...797...46S} type abundances within the shell, as given in Table \ref{tab:abdmodelsshen}. Of the three models considered, that with a helium envelope of 0.005\Msun{} fits the observed feature of SN~2011fe most closely. As this model does not include any calcium in the shell, the CaNIR feature is the result of only absorption within the ejecta and does not reproduce the high velocity wing of the feature. Despite the shell containing over 22 per cent silicon, there is no evidence of a HVF in the \ion{Si}{II} 6355\Ang{} or other silicon features. This is related in part to the lack of calcium HVF in the helium substrate (see Appendix \ref{MZW1:apdx:he}) --- the helium does not ionize easily, leading to lower electron density and higher ionization state of the silicon within the shell.

The 0.01\Msun{} envelope model is similar to that of the helium substrate model with 3~dex enhancement of calcium(not shown), with a CaNIR feature extending to about 6300\Ang{}; the 0.02\Msun{} envelope model generates a strong calcium feature between about 5000 -- 7500\Ang{}. These two models clearly can be ruled out at this epoch.

\subsection{5~days after the explosion}



\subsubsection{Solar-type with hydrogen substrate}



\begin{figure}\centering
\includegraphics[height=\columnwidth,angle=-90]{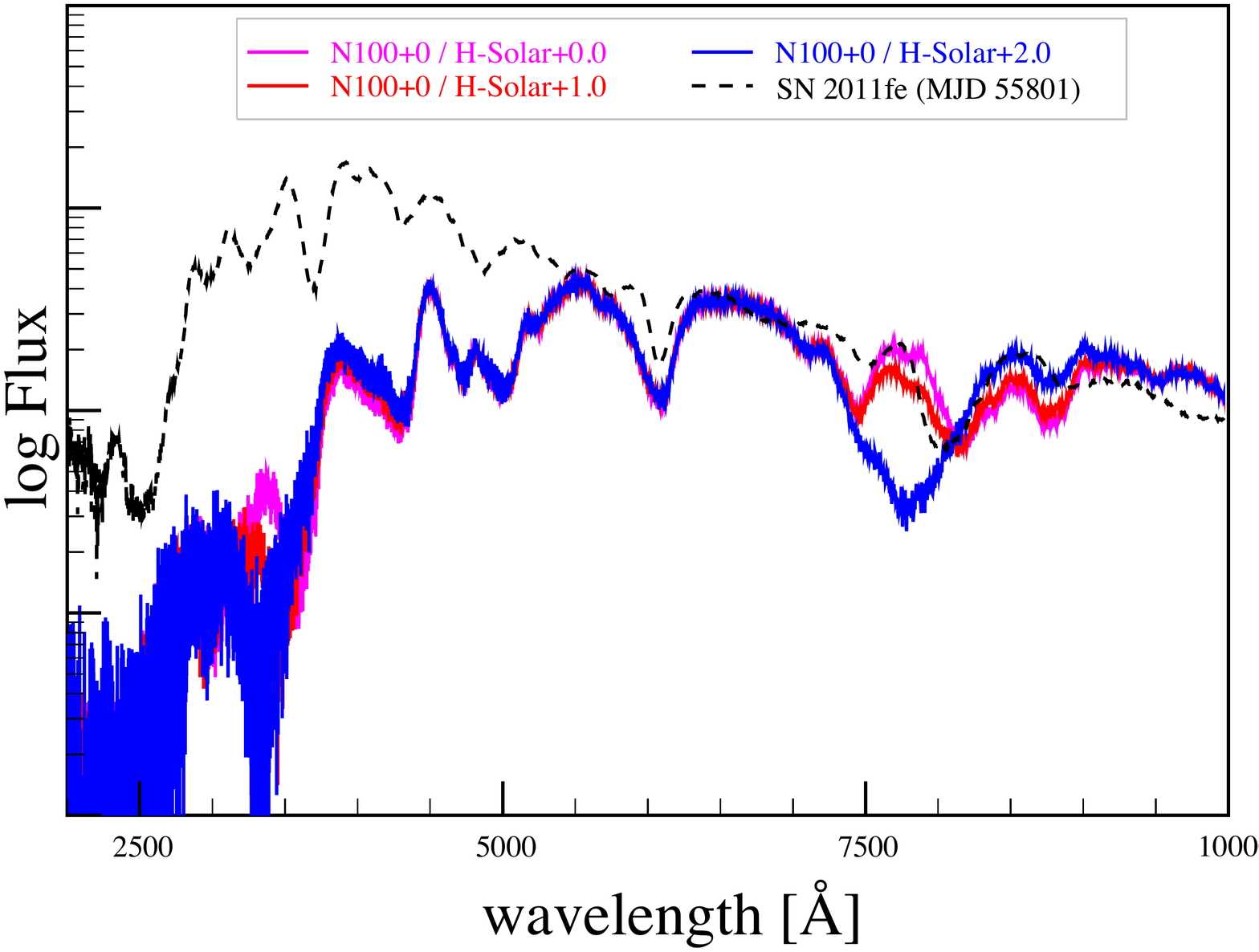}
\center
\includegraphics[height=\columnwidth,angle=-90]{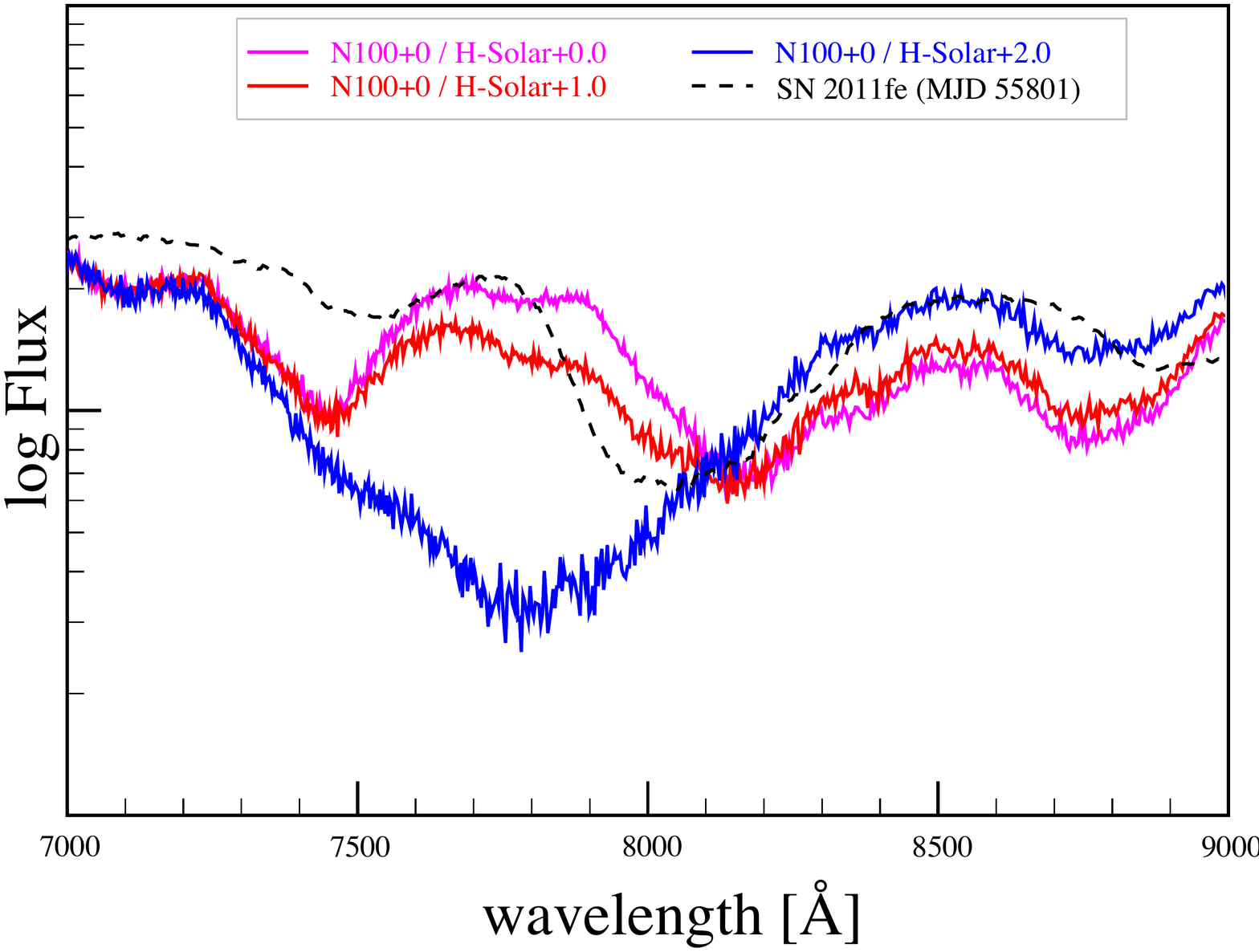}
\caption[H substrate shells at 5~d]{\label{fig:d5-H} Top: Like Figure \ref{fig:d2-H} but at 5~d after the explosion, and with calcium enhancement in the shell ranging from 0 -- +2~dex. At this epoch, the calcium in the shell continues to only affect the CaNIR feature.
Bottom: Like the top panel but focused on the CaNIR feature at 5~d after the explosion. For the shell with an enhancement of +1~dex of calcium, the flattening of the P~Cygni peak of the \ion{O}{I} 7773\Ang{} feature (near 7550\Ang{}) is the HVF. At this epoch, an enhancement by a factor of about 10 greater than the enhancement required at 2~d is necessary to generate a HVF. At 7800 \AA, the models from top to bottom represent N100+0/H-solar+0.0, N100+0/H-solar+1.0, and N100+0/H-solar+2.0, respectively.
A color figure is available online.}
\end{figure}

Figure \ref{fig:d5-H} shows the spectra that result from abundance model N100+0 for the ejecta and the hydrogen substrate abundance models for the shell at 5~d after the explosion. There continues to be no evidence of hydrogen absorption or emission within the synthetic spectra at this epoch, nor any effect of the calcium within the shell upon the \ion{Ca}{II} H\&K feature. 

The bottom panel of Figure \ref{fig:d5-H} focuses on the CaNIR feature. At this epoch, an enhancement of near 1~dex of calcium within the shell, relative to a solar abundance, is required to match the CaNIR HVF of SN~2011fe. Enhancement of calcium by 2~dex within the shell causes an excessively strong feature. The blue edge of the feature is greatly softened by the material within the shell. The enhancement of calcium within the shell required to generate a HVF at this epoch is a factor of about 10 greater than the enhancement required at 2~d; we discuss the meaning and implications of this in Sec. \ref{MZW1:ssec:discussion}.

The \ion{Si}{II} 6355\Ang{} and \ion{O}{I} 7773\Ang{} features resulting from the models are stronger than the observed features at this epoch, with the \ion{O}{I} feature being particularly enhanced. Both of these are associated with the ejecta rather than the shell and may represent an overabundance of silicon and oxygen at velocities higher than that of the photosphere within either the \citet{2005ApJ...623..337G} explosion model or an overabundance in the yields of oxygen and silicon within the \citet{2013MNRAS.429.1156S} N100 model.

\begin{figure}\centering
\includegraphics[height=\columnwidth,angle=-90]{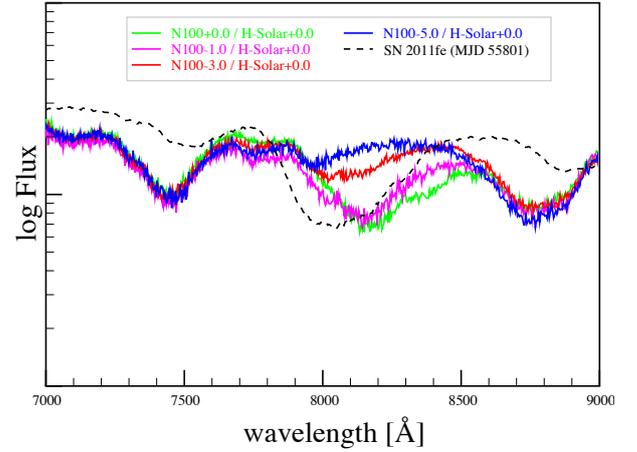}
\caption[Varying ejecta composition at 5~d (CaNIR)]{\label{fig:d5-Seit}
Like Figure \ref{fig:d5-H} at 5~d after the explosion but varying the calcium abundance in the ejecta. At 8200 \AA, the models from bottom to top represent N100+0/H-solar+0.0, N100-1/H-solar+0.0, N100-3/H-solar+0.0, and N100-5/H-solar+0.0, respectively. The calcium within the ejecta tends to affect only the appearance of the feature between 8000 -- 8500\Ang{}. A calcium depletion of -2 -- -1~dex relative to the \citet{2013MNRAS.429.1156S} N100 model with stable nuclides may be the best approximation of the observed feature for a shell with a solar abundance of calcium. It is notable that there is a slight knee near 8300\Ang{} that is visible in the observed feature as well as that associated with the \citet{2013MNRAS.429.1156S} abundance but is weaker or not apparent when the calcium is depleted. A color figure is available online. }
\end{figure}

Figure \ref{fig:d5-Seit} shows the effect of calcium within the ejecta on the synthetic spectra, depleting the calcium yield relative to the \citet{2013MNRAS.429.1156S} N100 model with stable nuclides by -5 -- 0~dex. The effect of the ejecta on the CaNIR feature is entirely redward of 7900\Ang{}, with the minimum occurring near 8150\Ang{}. There is a significant HVF between 7800 -- 8000\Ang{} that cannot be explained entirely by the ejecta at this epoch. It is again not clear that any one model of abundance in the ejecta results in the CaNIR feature better fitting the observed feature at this epoch, though a calcium yield within $-3$ -- 0~dex of the \citet{2013MNRAS.429.1156S} N100 model is likely necessary to explain the PVF.

\begin{figure}\centering
\includegraphics[height=\columnwidth,angle=-90]{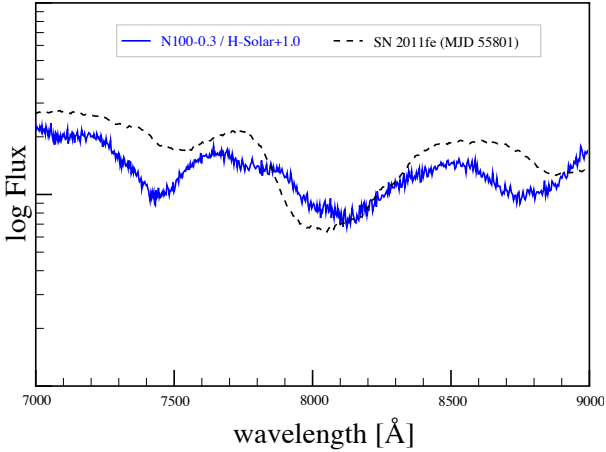}
\caption[Best H substrate at 5~d (CaNIR)]{\label{fig:d5-H-CaNIR-Best} 
Same as the bottom panel of \ref{fig:d5-H} at 5~d after the explosion, but using abundances that represents a combination of depletion of calcium in the ejecta and enhancement of calcium in shell that provides a better fit to the CaNIR feature. The abundance of calcium in the shell is a balance between attempting to achieve the observed depth between 7900 -- 8000\Ang{}, and the peak near 7700\Ang{}. The P~Cygni emission near 8500\Ang{} is slightly weaker than observed in SN~2011fe. A color figure is available online.}
\end{figure}

Figure \ref{fig:d5-H-CaNIR-Best} shows an effort to produce a best combination of ejecta and shell abundances to match the observed CaNIR feature of SN~2011fe. We select a depletion of 0.3~dex of calcium relative to the \citet{2013MNRAS.429.1156S} N100 model with stable nuclides, and 1.0~dex of enhancement of calcium within the shell, relative to a solar value. This model represents a balance between matching the depth of the feature blueward of 8000\Ang{}, while avoiding excessive absorption blueward of 7800\Ang{}. In this model, the P~Cygni emission is slightly too weak compared to the observed CaNIR feature.

\subsubsection{Shen \& Moore-type envelopes}

\begin{figure}\centering
\includegraphics[height=\columnwidth,angle=-90]{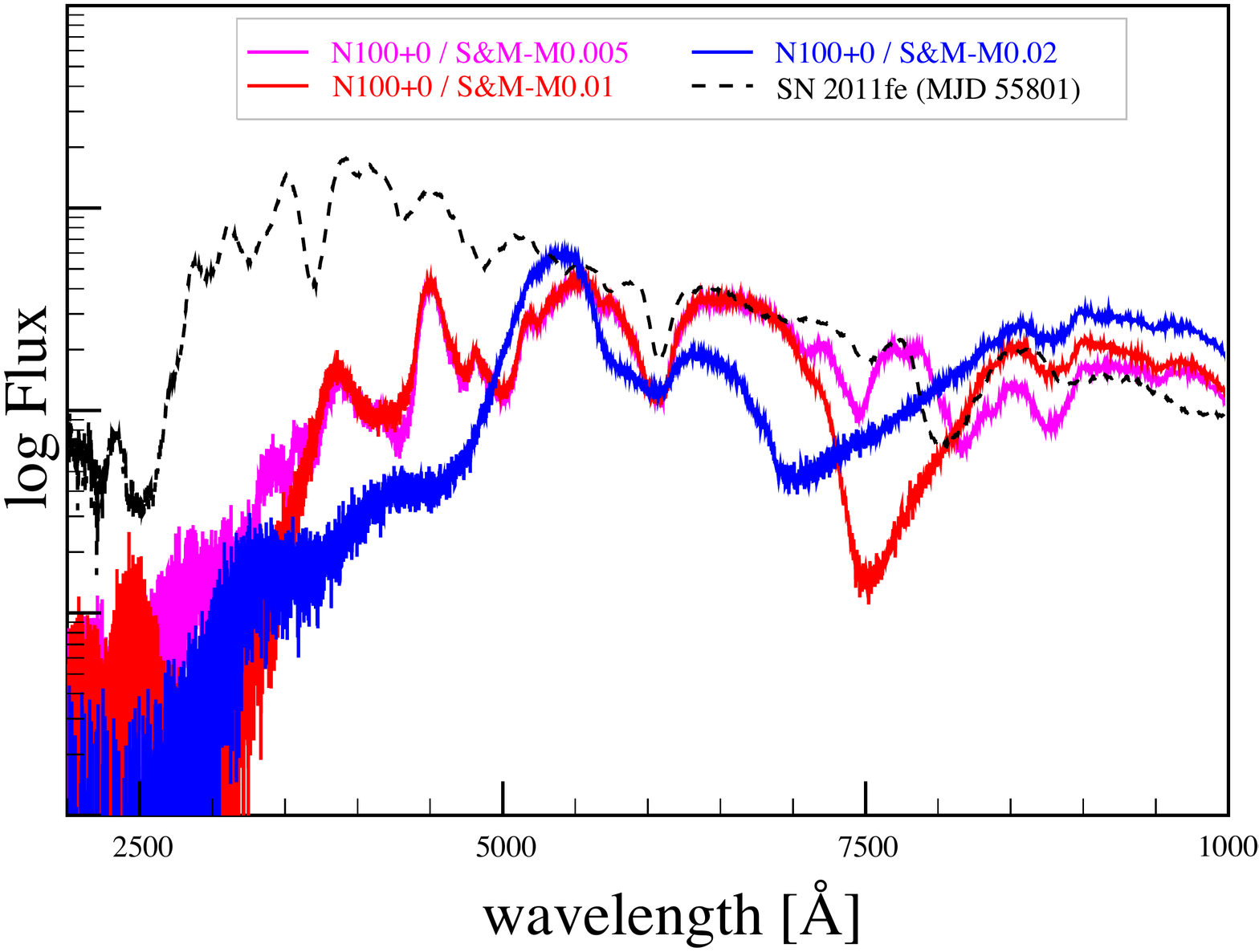}
\caption[Shen \& Moore envelopes at 5~d]{\label{fig:d5-Shen} Like Figure \ref{fig:d5-H} at 5~d after the explosion, but using \citet{2014ApJ...797...46S} type abundances for the shell. The model with a 0.005\Msun{} envelope is shown in magenta (uppermost spectrum at 7500 \AA), the 0.01\Msun{} envelope is shown in red (bottom spectrum at 7500 \AA), and the 0.02\Msun{} envelope is shown in blue (middle spectrum at 7500 \AA). The 0.01 and 0.02\Msun{} models result in an extremely deep and extended CaNIR feature that does not match any observed features. A color figure is available online. }
\end{figure}

Figure \ref{fig:d5-Shen} shows the synthetic spectra at 5~d for the \citet{2014ApJ...797...46S} type abundance models. The models with an envelope of mass 0.01\Msun{} and 0.02\Msun{} have an excessively strong and extended CaNIR feature; the 0.02\Msun{} model also shows evidence of an HVF for the \ion{Si}{II} 6355\Ang{} feature that is much stronger than those that are observed. The model with an envelope of mass 0.005\Msun{} is very similar to that of the models with a helium substrate for the shell but does not have any enhancement of calcium features due the absence of calcium in the envelope.

\subsection{9~days after the explosion}

At 9~d after the explosion, there appears a feature in the spectra of SN~2011fe at about 8240\Ang{} that is clearly the HVF. As you will see in the following subsections and figure, it is difficult to reproduce this feature with our \textsoft{tardis} models.


\subsubsection{Solar-type with hydrogen substrate}

\begin{figure}\centering
\includegraphics[height=\columnwidth,angle=-90]{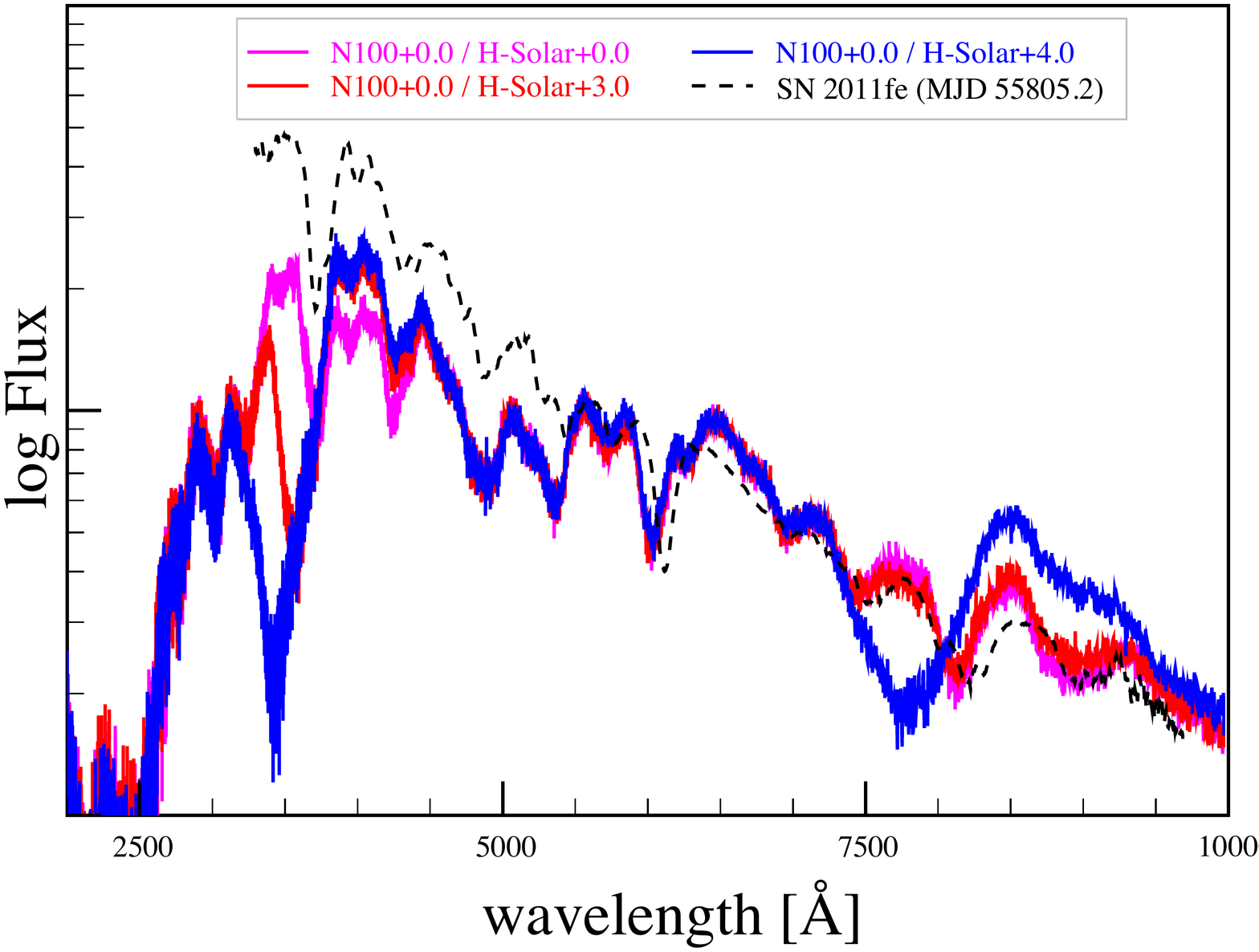}
\center
\includegraphics[height=\columnwidth,angle=-90]{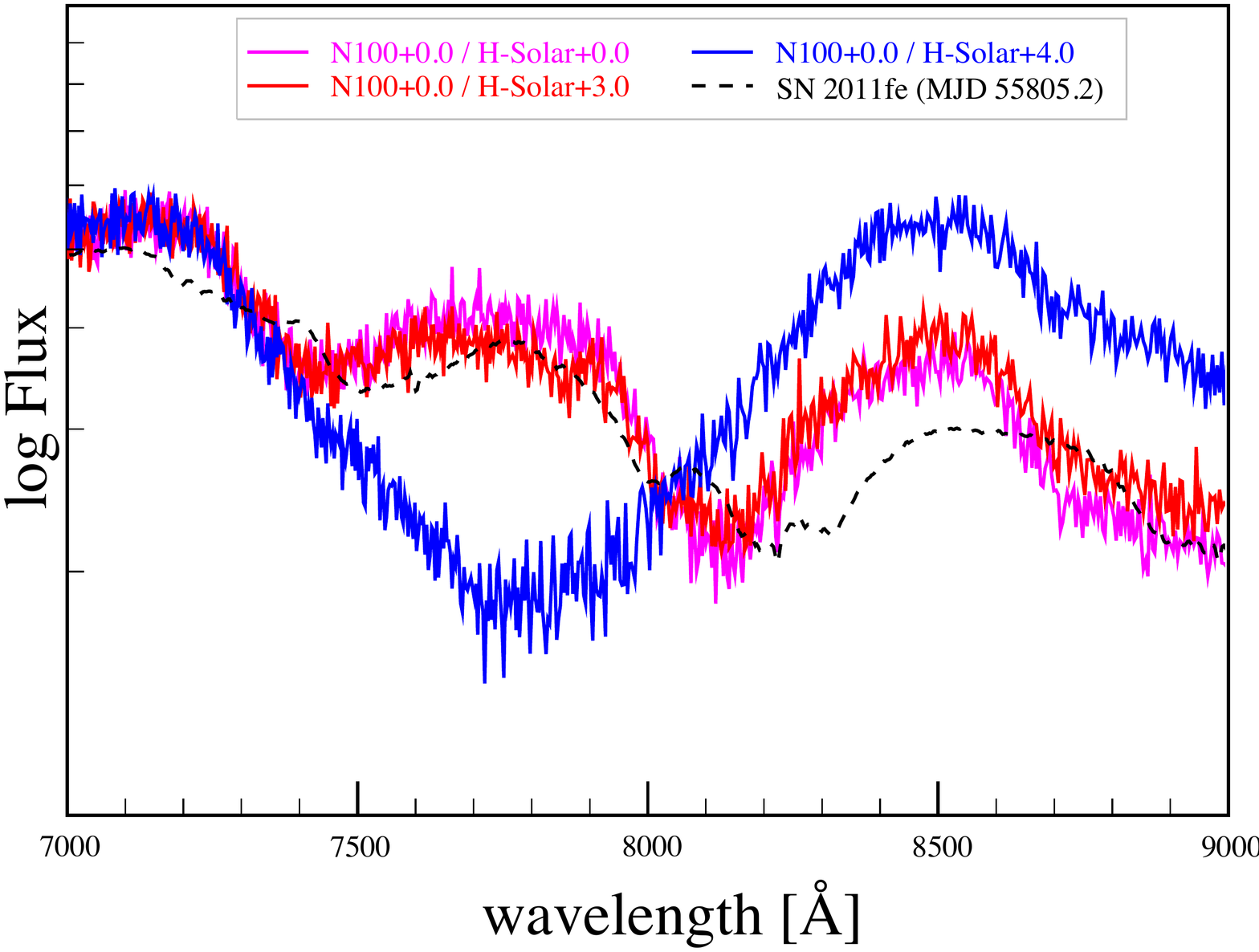}
\caption[H substrate shells at 9~d]{\label{fig:d9-H} Top: Like Figure \ref{fig:d2-H} but at 9~d after the explosion, and for a shell with calcium enhancement between 0 -- +4~dex. At this epoch, the calcium within the shell affects both the CaNIR and \ion{Ca}{II} H\&K features. There continues to be no evidence of absorption or emission due to the hydrogen in the shell.  
Bottom: Like the top panel but focused on the CaNIR feature. An enhancement of +4~dex results in a strong, broad CaNIR HVF, while there is little difference between the models with 0 -- +3~dex of calcium enhancement. At this epoch, the feature resulting from the models produces an acceptable fit to the the blue side of the CaNIR feature, but performs poorly near 8200\Ang{}.
At 7800 \AA, the models from top to bottom represent N100+0/H-solar+0.0, N100+0/H-solar+3.0, and N100+0/H-solar+4.0, respectively. A color figure is available online. }
\end{figure}

Figure \ref{fig:d9-H} shows the spectra that result from abundance model N100+0 for the ejecta and the hydrogen substrate abundance models for the shell at 9~d after the explosion. There continues to be no evidence of hydrogen absorption or emission within the synthetic spectra at this epoch, but the calcium within the shell does affect the \ion{Ca}{II} H\&K feature. The \ion{O}{I} 7773\Ang{} and \ion{Si}{II} 6355\Ang{} features better match the observed features in SN~2011fe, although both features have minima that are too blue by $\sim$100\Ang{}. We computed models with lower photospheric velocities, but the overall nature of the resulting model spectra deteriorated substantially, characterized by substantial deficit in the blue. The low velocities of the \ion{O}{I} 7773\Ang{} and \ion{Si}{II} 6355\Ang{} features may be a structural problem with the models of \citet{2005ApJ...623..337G} rather than a problem with the photosphere velocity.

\begin{figure}\centering
\includegraphics[height=\columnwidth,angle=-90]{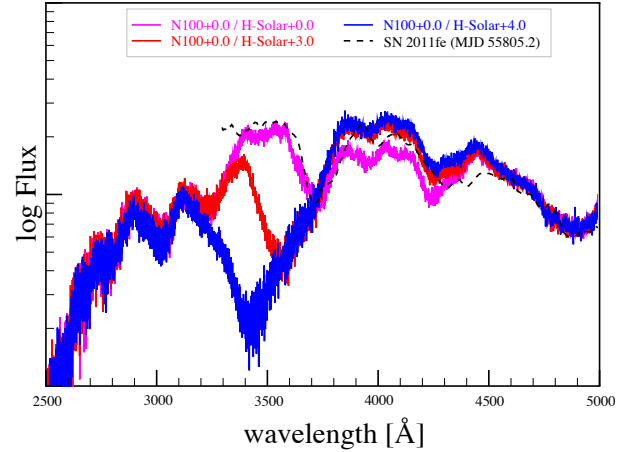}
\caption[H substrate shells at 9~d (Ca H\&K)]{\label{fig:d9-H-CaHK} Like 
Figure \ref{fig:d9-H} at 9~d after the explosion, but focused on the \ion{Ca}{II} H \& K feature. At 3500 \AA, the models from top to bottom represent N100+0/H-solar+0.0, N100+0/H-solar+3.0, and N100+0/H-solar+4.0, respectively. A calcium enhancement of +1~dex or more above solar value for material in the shell affects the strength of the feature and appears as a HVF. A calcium abundance near solar provides a better fit to this feature at this epoch. A color figure is available online. }
\end{figure}

The bottom panel of Figure \ref{fig:d9-H} focuses on the CaNIR feature. Up to +3~dex of enhancement of calcium above solar value within the shell has little to no effect on the CaNIR feature. An enhancement of +4~dex produces a strong HVF between 7500 -- 8000\Ang{}. At this epoch, the calcium within the ejecta results in a feature that is too blue by about 200\Ang{}, or at least is lacking absorption near 8300\Ang{}. Figure \ref{fig:d9-H-CaHK} shows the \ion{Ca}{II} H\&K feature at this epoch. Enhancement of calcium within the shell leads to an overly deep and extended feature; this provides a constraint on the degree to which calcium can be enhanced in the shell without producing an overly strong \ion{Ca}{II} H\&K feature.

\begin{figure}\centering
\includegraphics[height=\columnwidth,angle=-90]{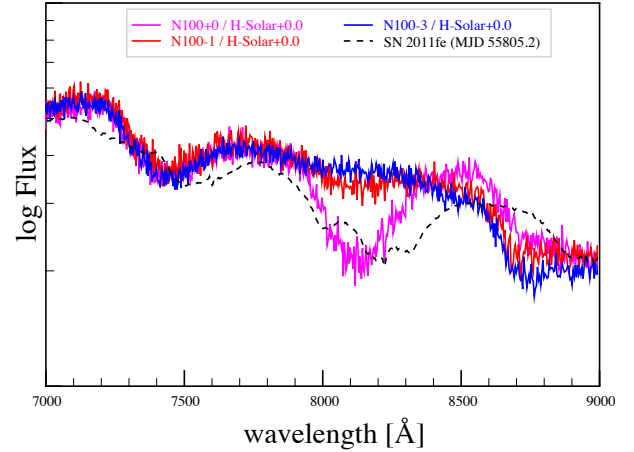}
\caption[Varying ejecta composition at 9~d (CaNIR)]{\label{fig:d9-Seit} Like Figure 
\ref{fig:d9-H} at 9~d after the explosion, but showing variation in the calcium abundance in the ejecta. At 8200 \AA, the models from bottom to top represent N100+0/H-solar+0.0, N100-1/H-solar+0.0, and N100-3/H-solar+0.0, respectively. A reduction of -1~dex of calcium within the ejecta notably reduces the strength of the PVF at this epoch and does not do a good job of recreating the feature. An abundance of calcium that is near that of the \citet{2013MNRAS.429.1156S} N100 model with stable nuclides is likely the best match at this epoch. A color figure is available online. }
\end{figure}

Figure \ref{fig:d9-Seit} shows the effect of calcium within the ejecta on the synthetic spectra, depleting the calcium yield relative to the \citet{2013MNRAS.429.1156S} N100 model with stable nuclides from  -5 -- 0~dex. The CaNIR feature becomes noticeably weaker with even a modest depletion of calcium in the ejecta. At this epoch, the observed feature in SN~2011fe is better matched by little to no depletion of calcium within the ejecta.

\begin{figure}\centering
\includegraphics[height=\columnwidth,angle=-90]{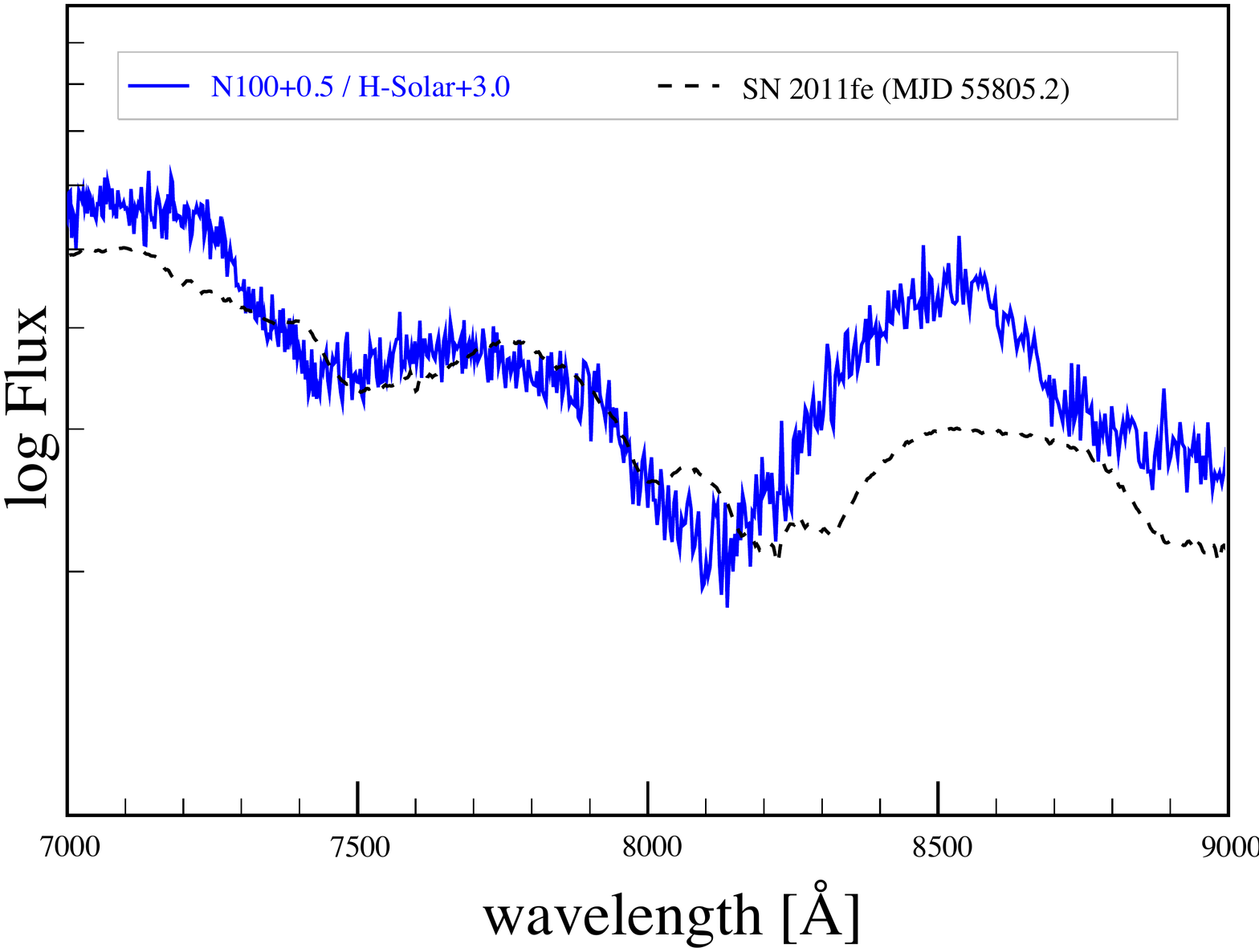}
\caption[Best H substrate at 9~d (CaNIR)]{\label{fig:d9-H-CaNIR-Best} Same as Figure 
\ref{fig:d9-H} at 9~d after the explosion, but using abundances that represent a combination of enhancement of +0.5~dex of calcium in the ejecta and +3~dex in the shell that provides a better fit to the CaNIR feature. At this epoch this model agrees in the blue wing, but fails to capture the absorption on the redward side of the feature. We were unable to reproduce the apparent multiple component nature of the feature at this epoch. The multiple components of the PVF may be a result of variations in calcium abundance within the ejecta not captured by the explosion models that we have used. A color figure is available online. }
\end{figure}

Figure \ref{fig:d9-H-CaNIR-Best} shows an effort to produce a best combination of ejecta and shell abundances to match the observed CaNIR feature of SN~2011fe. We select an enhancement of +0.5~dex of calcium in the ejecta, relative to the \citet{2013MNRAS.429.1156S} N100 model with stable nuclides, and +3.0~dex of enhancement of calcium within the shell, relative to a solar value. This model results in slightly too much absorption near 8100\Ang{}, due to absorption within the ejecta, but matches the blue wing of the feature reasonably well. The model produces excessive flux on the red wing. Even with our best fitting model, we are unable to recreate the three components of the CaNIR feature seen in SN~2011fe at this epoch.

\subsubsection{Shen \& Moore-type envelopes}

\begin{figure}\centering
\includegraphics[height=\columnwidth,angle=-90]{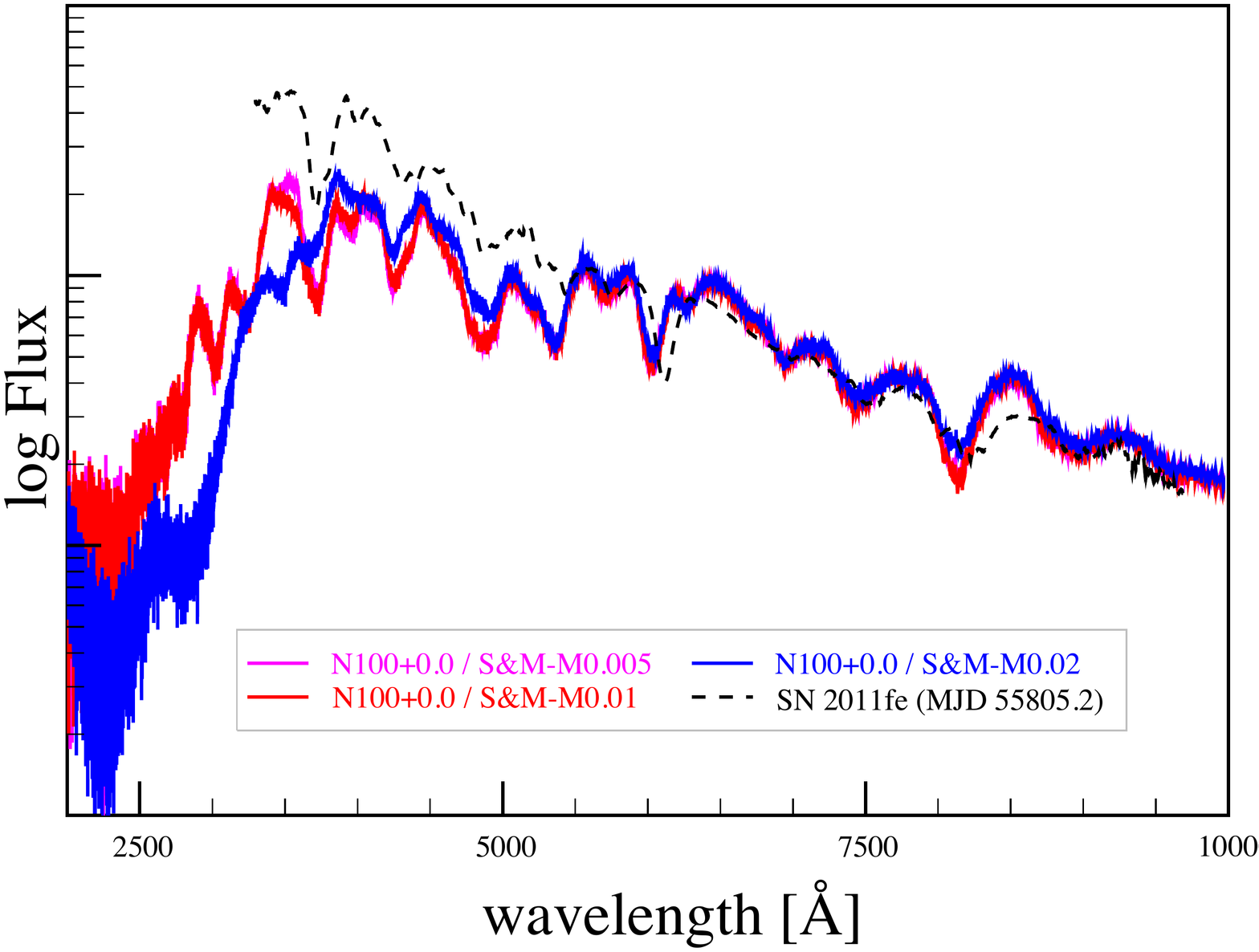}
\caption[Shen \& Moore envelopes at 5~d]{\label{fig:d9-Shen} Like Figure \ref{fig:d9-H} at 9~d after the explosion, but with \citet{2014ApJ...797...46S} type abundances for the shell. The model with a 0.005\Msun{} envelope is shown in magenta (uppermost spectrum at 3500 \AA), the 0.01\Msun{} envelope is shown in red (middle spectrum at 7500 \AA), and the 0.02\Msun{} envelope is shown in blue (bottom spectrum at 7500 \AA). The three models are nearly identical longward of 4000 \AA. The 0.01 and 0.005\Msun{} models are nearly indistinguishable at this epoch, with the exception of some additional absorption by the 0.01\Msun{} envelope on the blue side of the \ion{Ca}{II} H\&K feature. The 0.02\Msun{} model results in a weaker CaNIR feature as well as weaker features blueward of $\sim$5000\Ang{}, in addition to continued depletion of flux blueward of $\sim$4000\Ang{}. A color figure is available online. }
\end{figure}

Figure \ref{fig:d9-Shen} shows the synthetic spectra at 9~d for the \citet{2014ApJ...797...46S} abundance models. At this epoch, the spectra generated from the 0.005\Msun{} and 0.01\Msun{} envelope models match that of the solar-type abundance models. The 0.02\Msun{} model results in the flux blueward of 3500\Ang{} to be affected, chiefly depletion by the iron present in the shell, and making the \ion{Ca}{II} H\&K feature indistinct due to blending with absorption bands in the near-UV.

\subsection{Discussion}\label{MZW1:ssec:discussion}

In the prior subsections, we have considered shells with a mass of 0.005\Msun{} with a hydrogen substrate and varying levels of enhancement of depletion of calcium within the shell or ejecta. Models with helium, or carbon-oxygen substrates are considered in the appendices. We find that the substrate has no effect on the spectrum, other than a higher abundance of calcium required in shells with a helium substrate due to the higher ionization potential (and hence lower electron density) of helium. Table \ref{bestfit} summarizes the preferred calcium abundances in the shell and in the ejecta at the three epochs we have considered 2, 5 and 7 days in comparison with SN~2011fe as given for Figures \ref{fig:d2-H-CaNIR-Best}, \ref{fig:d5-H-CaNIR-Best}, and \ref{fig:d9-H-CaNIR-Best}.
We also considered shells with masses of 0.005, 0.01, and 0.02\Msun{} and employing related abundances of the helium envelopes of \citet{2014ApJ...797...46S}. The envelopes with a mass of 0.005\Msun{} do not contain calcium, so fail to produce an HVF; the envelopes of 0.01 and 0.02\Msun{} result in overly broad and strong HVF.

\begin{table}
\caption{Calcium Abundance Parameters in Preferred Models with Hydrogen Shell Substrate}
\label{bestfit}
\begin{center}
\begin{tabular}{lcc}
\hline
Epoch  & Shell$^1$ & Ejecta$^2$   \\

\hline

2 days & +0.4 & -2.0  \\
\\
5 days & +1.0  & -0.3 \\  
\\                                                                    
9 days & +3.0  & +0.5  \\

\hline  

\end{tabular}
\end{center}

$^1$ dex with respect to solar.\\
$^2$ dex with respect to model N100 of \citet{2013MNRAS.429.1156S}.

\end{table}

Regarding shells with a mass of 0.005\Msun{} with a hydrogen substrate, \citet{2013ApJ...762L...5S} and \citet{2018ApJ...852L...6B} find upper limits of $10^{-3}$\Msun{} and $10^{-4}$\Msun{}, respectively, of hydrogen during the nebular phase of SN~2011fe. These limits are based on the assumption of the hydrogen being stripped from a companion and residing at relatively low velocities ($\sim$1,000--2,000\kmSec{}) within the ejecta. It is not clear that this limit also applies to hydrogen that lies at the leading edge of the ejecta, well separated from the excitational radiation available from the nickel and cobalt decay nearer the centre of the supernova. For some models, we have extended the \textsoft{tardis} results into the far-infrared. For our solar-type abundance models, there are no observable absorption or emission features due to the substrate within the near-ultraviolet to far-infrared range at the epochs that we consider.

A common result for all composition models at 2 and 5~d is that the PVF is too strong using the \citet{2013MNRAS.429.1156S} N100 composition model. There are many possible reasons for this: the abundance of calcium as a function of velocity within the ejecta may not be correct due to the combination of elements of similar atomic mass into groups in the explosion models of \citet{2005ApJ...623..337G}; the distribution of silicon group elements in the explosion models may not represent the real distribution of these elements within SN~Ia; the \citet{2013MNRAS.429.1156S} N100 explosion model may have an overabundance of calcium relative to other silicon group elements, or the physics that is incorporated into \textsoft{tardis} does not fully explain the state of calcium within the ejecta. All of these factors likely play some role in affecting the appearance of the feature. 

We have not found a consistent abundance of calcium within the shell that can explain the observed HVF. In \citetalias{2018MNRAS.476.1299M}, we noted that the evolution of the parameters of the fit were inconsistent with the HVF fading solely due to the reduced density as the shell expands adiabatically. \citetalias{2018MNRAS.476.1299M} suggested that the ionization state of calcium within the shell (or ejecta) is likely not constant over the first few weeks after the explosion. While we expect that \textsoft{tardis} does more accurately determine the ion state of calcium, it still relies on an incomplete approximation of the ionization state of all elements. The varying apparent abundance of calcium that is required may be the result of this approximation. We also note that we have assumed a uniform composition of the shell throughout; if the shell material is not homogeneous there may be less calcium required than we have found here. A non-homogeneous distribution of calcium within the shell may also explain why the blue side of the CaNIR feature for our best fit model tends to have a shallower slope than is observed in SN~2011fe. 

At 9~d after the explosion, the CaNIR feature is better explained by only the calcium within the ejecta, although the PVF resulting from our models tends to be too shallow, slightly blue, and does not show the three components of the feature seen in SN~2011fe. The lack of a model HVF at this epoch is consistent with the findings of \citetalias{2018MNRAS.476.1299M}. \citetalias{2017MNRAS.467..778M} suggest that the strength of the PVF would overtake that of the HVF at about 8~d after the explosion, though we note that the photosphere velocity used in \citetalias{2017MNRAS.467..778M} is much faster (by 5,000 -- 14,000\kmSec{}) than the velocities used in this work, leading to a much weaker PVF in \citetalias{2017MNRAS.467..778M} at these early epochs. While \citet{2014MNRAS.437..338C} and \citet{2015MNRAS.451.1973S} both report a reasonably strong HVF component much later than 9~d after the explosion for SN~2011fe, we note that these works rely on fitting of multiple Gaussian features to an inherently non-Gaussian absorption feature. This fitting technique does not reliably capture the velocity or strength of the HVF once it blends with the PVF; see \citetalias{2017MNRAS.467..778M} sec. 4.1 for further details.

\subsubsection{Implications of super-solar calcium}
In each of the solar-type compositions that we consider at 2 and 5~d, we need a super-solar abundance of calcium in the shell to generate the HVF. While this may be due to incorrect assumptions within \textsoft{tardis} regarding the ionization or excitation state of calcium, it may accurately reflect the abundance of calcium in the shell. We can speculate on the source of this calcium by considering the possible sources of material for the shell.

If the source of the shell material is a companion star donating material, a super-solar calcium content may be suggestive of a relatively young companion. The ubiquity of HVF in SN~Ia and the expectation that there are likely both long-delay and short-delay SN~Ia suggest that many should have older companions with sub-solar abundances of calcium. Post-main sequence stars with masses below $\sim$7\Msun{} tend to be depleted in calcium within their atmospheres \citep{2016ApJ...825...26K, 2016ApJS..225...24P}, so these are less likely to be the source of such calcium.

If the donor star were instead a white dwarf, the average calcium abundance may be solar, but settling of metals toward the center of the donor white dwarf would deplete the calcium in the outer layers and enhance it within the inner layers \citep{1958whdw.book.....S}. The details of the merger process then dictate the abundance of calcium within the shell: it could be depleted in calcium, if it forms from the outermost layers of the donor; enhanced in calcium, if it forms from material nearer the core of the companion; or near the original metallicity of the star, if the material of the companion is mixed during the merger or accretion process. The original metallicity of the system should determine the strength of the HVF, suggesting that any SN~Ia that originate from population II systems should have weaker calcium HVF. The ubiquity of the CaNIR HVF again argues against this possibility.

Another implication of the calcium content of the shell material being related to the original metallicity of the system suggests that HVF should be weaker with increasing redshift. The comprehensive studies of HVF in SN~Ia that have been done to date \citep{2014MNRAS.437..338C, 2014MNRAS.444.3258M, 2015MNRAS.451.1973S} consider only nearby redshifts and no attempts have been made to evaluate the trend of the strength of the HVF with the redshift of the host galaxy.

Given that the HVF are common, generally similar in strength, and that variations of less than 1~dex in the abundance of calcium within the shell do not have significant effect on the HVF at 5~d and 9~d after the explosion (as well as at 2~d in some cases), it is possible that the calcium that generates the HVF is newly synthesized. This could be explained by material ejected from the supernova at high velocity, by nucelosynthesis on the surface of the progenitor during pre-explosion accretion, or by mixing of deeper layers of the supernova with the material in the shell. Our methodology here may be partially capturing that information, although it is unlikely that such ejected material would match our models in terms of density or kinematics. 

One mechanism by which freshly synthesized calcium could be introduced into the shell would be through the Rayleigh-Taylor instabilities that would occur during the interaction between the shell and ejecta. If this were the source of the enhanced calcium in the shell, however, there would be a relatively smooth gradient of calcium abundance from the ejecta to the shell material, resulting in an indistinct HVF, unless only the outermost layers of the ejecta are somehow enhanced in calcium due to or prior to the explosion. As the \citetalias{2017MNRAS.467..778M} models were produced through 1-D simulations, they do not include such mixing.

The helium envelopes of \citet{2014ApJ...797...46S} can also contain freshly-formed calcium. We considered envelopes with a mass of 0.005, 0.01, and 0.02\Msun{}, none of which produce spectra that are a good match to SN~2011fe or any other SN~Ia. We note that the 0.01 and 0.02\Msun{} envelopes generate overly deep and extended calcium features; such features require correct handling of the Lorentzian wings and are not well approximated by \textsoft{tardis}, but given that the spectra generated by these models are so different than observed spectra we rule these models out as possible explanations of SN~Ia. An envelope with a mass near 0.008\Msun{} or 0.09\Msun{} may have the right abundance of calcium after the detonation to generate the observed HVF, though an envelope with a mass of 0.09\Msun{} is likely excluded, as shells with such a high mass are likely to have a photosphere within the envelope at early epochs and result in an HVF that is too slow at later epochs. The calcium abundance as a function of the mass near 0.008\Msun{} has a very steep slope of calcium abundance as a function of mass. While the models of \citetalias{2017MNRAS.467..778M} and \citet{2014ApJ...797...46S} are dynamically inconsistent, we take this an an indication that the mass of the envelope must be very fine-tuned to generate a calcium HVF that is comparable to those that have been observed. It is not clear why an envelope with such a finely-tuned mass would be favoured for that in which the detonation occurs. If this model is correct, identification of the ratio of calcium:helium required to generate the calcium HVF will provide a precise estimate of the mass of the envelope.

\section{Conclusion}\label{MZW1:sec:conclusion}
We have presented synthetic spectra at 2, 5, and 9~d after the explosion of a Type~Ia supernova, generated using the \citet{2017MNRAS.467..778M} models of interaction between the supernova and a compact circumstellar shell using \textsoft{tardis} \citep{2014MNRAS.440..387K}. We apply abundance models to the shells that consist of either a hydrogen, helium, or carbon-oxygen substrate with overall solar abundance of metals and depletion or enhancement in the calcium abundance, as well as abundance models matching the yields of the \citet{2014ApJ...797...46S} helium envelope detonation models for envelopes with a mass of 0.005, 0.01, and 0.02\Msun{}. We use a \citet{2013MNRAS.429.1156S} model N100 stable nuclide yield for abundances within the ejecta, with variations in the calcium abundance to adjust for using an explosion model (\citet{2005ApJ...623..337G} model `c') that groups nucleosynthetic products into five groups (carbon, oxygen, magnesium, silicon, and iron).

We find that solar to super-solar abundances of calcium may be required to generate a high-velocity feature in the calcium near-infrared triplet for models involving the hydrogen, helium, and carbon-oxygen substrates for the material within the shell. In all cases, the substrate leaves no imprint upon the spectra, although the models with a helium substrate require a greater enhancement of calcium than do the models with a hydrogen or carbon-oxygen. The apparent need for super-solar abundance of calcium may be in tension with the ubiquity of high-velocity features, though such an enhanced abundance may come about due to synthesis of calcium as part of the explosion, or by the concentration of calcium, e.g, via settling, or pre-explosion surface nucleosynthesis, prior to the explosion. Accretion of hydrogen or helium from a non-degenerate companion can lead to shell flashes on the surface of the progenitor white dwarf. \citet{2018ApJ...863..125K} present a recent study of this possibility as well as a review of previous modeling of the phenomena. \citet{2018ApJ...863..125K} conclude that accretion-fed shell flashes could build up a layer enhanced in silicon surmounted by a layer enhanced in calcium especially in the case of a white dwarf with mass very close to the Chandrasekhar limit. They suggest that expulsion of this freshly synthesized surface material could be the source of HVF, with the calcium HVF typically moving somewhat faster than the silicon HVF. If it could be established that this process were responsible for the common presence of HVF in SN~Ia, it would point to a single degenerate progenitor evolution and a delayed-detonation explosion. The explanation for the apparent calcium enhancement may have more plebian explanations. Within our models, the super-solar abundance could also be a result of our assumption of the homogeneity of calcium within the shell, or approximations made within \textsoft{tardis} regarding the ion- or excitation state of calcium.

The \citet{2014ApJ...797...46S} models are extremely sensitive to the mass of the envelope in terms of generating spectra that are similar to those observed in SN~2011fe at equivalent epochs. An envelope of mass 0.005\Msun{} does not contain calcium and therefore does not produce an HVF, but neither does it produce a \ion{Si}{II} HVF despite containing 22 per cent silicon. An envelope of mass 0.01\Msun{} produces an excessively deep and extended (both in wavelength and temporally) CaNIR absorption feature. Our methods suggest than an envelope with a mass near 0.008\Msun{} would likely produce a HVF that is similar to those that have been observed; however, the models considered in \citet{2017MNRAS.467..778M} and \citet{2014ApJ...797...46S} are not consistent so we take this only as indication that the mass of the envelope must be finely-tuned rather than having this precise mass. In the lower mass envelopes, the helium:calcium ratio is extremely sensitive to the mass of the envelope, so the ratio that is required to generate the observed HVF would be useful to constrain the mass of the envelope should the helium-envelope detonation model be found to be the cause of SN~Ia. 

We have clearly not solved the question of the nature of the HVF and their common occurrence in SN~Ia but have put many of the issues in a new context by computing detailed spectral models. As in \citetalias{2017MNRAS.467..778M} and \citetalias{2018MNRAS.476.1299M}, we emphasize that a successful model of the HVF must account for its evolution in time. Within the context of the shell model, it is somewhat difficult to precisely reproduce the strongly blended profile of the CaNIR line at 2 and 5~d after the explosion. A reasonable representation can be achieved by judicious choice of the calcium abundance in the ejecta and the shell. The precise shape of the feature may depend on details like departures from spherical symmetry that we do not address. We were unable to reproduce the spectral details at 9~d after the explosion, especially the distinct feature that corresponds to what is easily identified as the separate HVF at that epoch. 

An initial primary motivation of this work was to try to identify the composition of the substrate in which the shell calcium resides. This would give especially important constraints on the progenitor system and the explosion mechanism. If the shell model is roughly correct, our models show that it is very difficult to determine the composition of the substrate of the high-velocity shell that carries the observable calcium. While there is no direct evidence for or against a helium substrate for the shell, our models suggest that a ``helium-rich" shell needs to be composed of mostly calcium rather than helium at 9~d after the explosion to generate a HVF, with a caveat that this demand for a large mass fraction of calcium may be an artifact of the ionization and excitation physics in \textsoft{tardis}. If we take this result at face value, it is strongly suggestive that the material responsible for the calcium HVF is not helium rich.

An important issue to consider is whether there are other, weaker features that appear in the vicinity of the CaNIR line. Some of our models display a small dip around 7800\Ang{}. This feature is probably a weaker line of some element in the ejecta and hence associated with the PVF, but it appears in the rough vicinity of the HVF of CaNIR. This feature could also be a reflection of the particular structure of the calcium abundance in our models. In any case, analysis of the HVF of CaNIR calls for caution in regard to possible contamination by other lines. 

The possibility of contaminating lines raises another, more radical possibility, the issue of whether the HVF are truly real. Could the HVF be evidence of other, weaker lines rather than evidence of calcium at high velocities? The easiest way to explain why HVF track the PVF but with a roughly constant displacement to higher velocity is that the features associated with the HVF are actually PVF. If this were the case for the CaNIR feature, what of other species in which HVF have been identified, e.g. silicon, oxygen, and iron? Could those also be weaker mis-identified  photospheric features? If so, what are the implications for compositions, progenitors, and explosion dynamics?

Much more work is called for to understand the ubiquitous HVF. Other supernovae should be studied in depth as we do here for three epochs of data on SN~2011fe. Not all events that display HVF manifest them in exactly the same way as does SN~2011fe. Because they are known to be polarized, HVF need to be studied with 3D dynamical and radiative transfer models. More care should be given to the details of the ionization and excitation of calcium. Another issue is whether the HVF earlier than 8~d should be treated differently than those at later epochs. At the earlier epochs, the HVF may be blended with the PVF, so it can be challenging to distinctly identify each component. Our shell models give broadening that is distinctly non-Gaussian, so identifying HVF by Gaussian fitting remains questionable.

We have concentrated on shell models as a plausible place to begin. Other models could be explored in the same depth we have tried to bring, stressing that any successful model must reproduce the line profile and the evolution of the HVF and PVF with time. Progenitor models must account for the ubiquity of the HVF and should also account for the absence in 91bg-like events. The issue of whether 91bg-like events are or are not related to ``typical'' SN~Ia and whether they can or cannot be produced with Chandrasekhar-mass models is of great interest. The fact that 91bg-like supernovae do not show HVF (the dog that did not bark in the night) \citep{2015MNRAS.451.1973S} may be an important clue that they are substantially different in some manner. 

Another issue of great interest is the growing number of SN~Ia that, like SN~2011fe, have data at very early times. \citet{2018arXiv180707576S} and \citet{2018arXiv180806343J} give summaries. Among SN~Ia that otherwise are similar near and after peak light, some, like SN~2011fe, show an early rise scaling about as $t^2$ \citep[but see][]{2017MNRAS.472.2787N}, early red colors, and relatively weak \ion{C}{II}; others show an early rise that is roughly linear in luminosity, early nearly constant blue colors, strong \ion{C}{II}, somewhat larger peak luminosity, slower declines, and are classified as 91T-like or Shallow Silicon in the characterization of \citet{2006PASP..118..560B}. Examples of this latter behavior are SN~2017cbv \citep{2017ApJ...845L..11H} and SN~iPTF2016abc \citep{2018ApJ...852..100M}. 
Various mechanisms are suggested to account for this bi-modal behavior including interaction with a companion and nickel mixing, but a commonly considered possibility is interaction with nearby circumstellar material, the basis for our current investigation of HVF. While there is no firm evidence in favor of a given model, it is clear that the red/blue bifurcation in early properties could have some implication for understanding HVF and vice versa. 

Given the ubiquity of the HVF, they should appear in both types of SN~Ia, and they do \citep{2015MNRAS.451.1973S}, although SN~iPTF2016abc may be an exception \citep{2018ApJ...852..100M}. It cannot be so simple that the early blue events represent some sort of circumstellar interaction that produces or suppresses HVF. Neither is it clear that outward mixing of nickel could account for HVF in a straightforward way. It could be that ejecta/shell interaction is the common mechanism for HVF with compact shells, as considered here, not contributing significantly to early light and somewhat more extended shells contributing early blue light. On the contrary, it is difficult to see how nickel mixing could contribute to the red/blue bifurcation and still lead to ubiquitous HVF whether the mixing is nil or extensive. 

As pointed out by \citet{2018ApJ...852..100M}, pulsating delayed detonation models \citep{1991A&A...245L..25K} might provide a natural way to produce some nearby unbound material that could play a role in forming the HVF. Shell-burning pulsations on the surface of the progenitor \citep{2018ApJ...863..125K} could generate HVF through explicit high-velocity composition structures, but it is unclear how this phenomenon would relate to the red/blue bifurcation. Consideration of ejecta/shell interaction still seems a fruitful path to understand HVF and perhaps the early red/blue distribution. The origin of appropriate shells should give crucial new understanding to the progenitor evolution and explosion mechanism of SN~Ia.

\appendix

\section{Solar-type abundance with helium substrate}\label{MZW1:apdx:he}

\subsection{2 days}

\begin{figure}\centering
\includegraphics[height=\columnwidth,angle=-90]{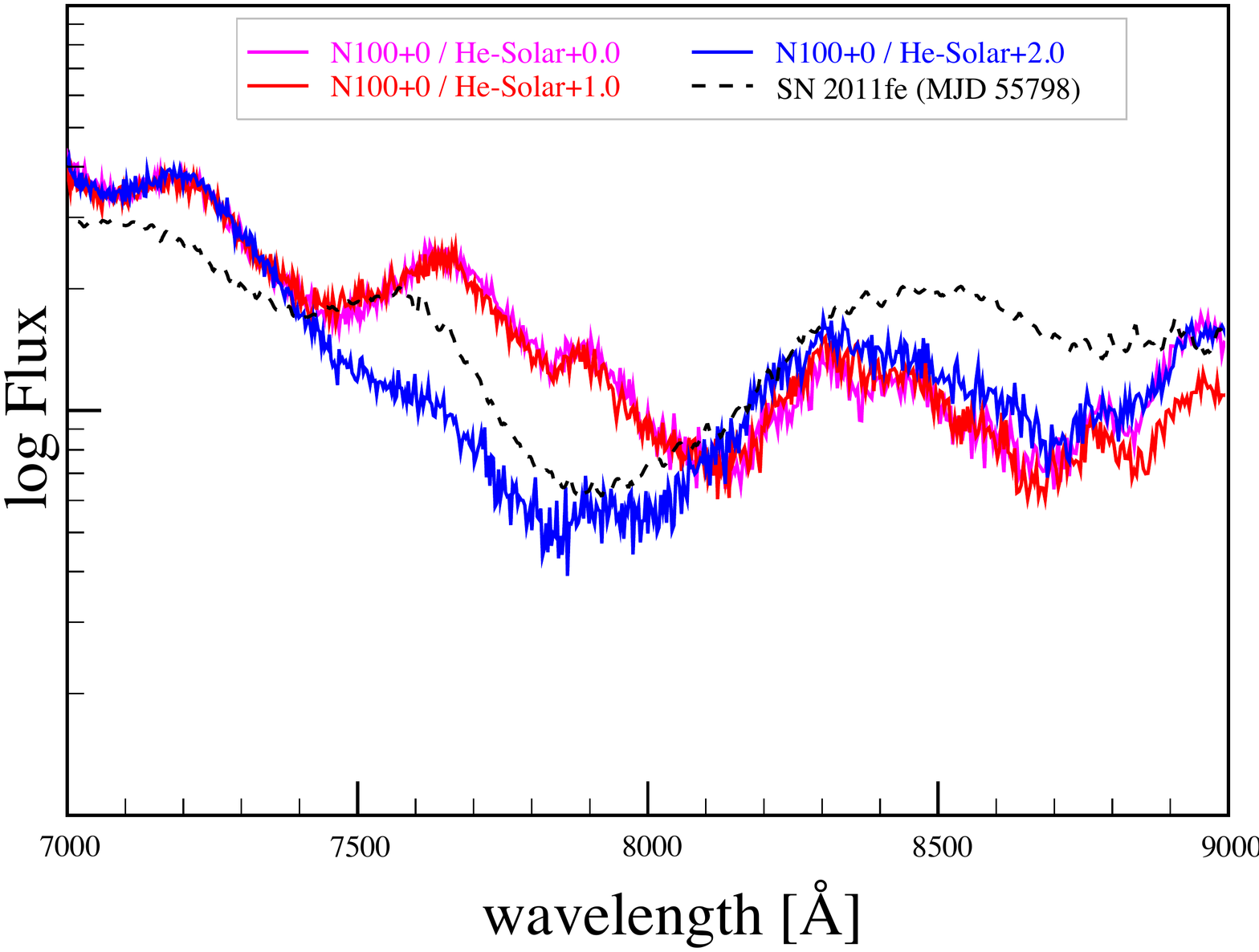}
\caption[He substrate shells at 2~d (CaNIR)]{\label{fig:d2-He} Like Figure 
\ref{fig:d2-H} at 2~d after the explosion, but for a shell with a helium substrate and calcium enhancement in the shell ranging from 0 -- +2~dex.
The models corresponding to N100+0/He-solar+0.0 and N100+0/He-solar+1.0 are nearly identical throughout. At 7800 \AA, these models are the uppermost and N100+0/He-solar+2.0 is the lower model. At 0 -- +1~dex enhancement above a solar abundance of calcium, there is little to no effect on the CaNIR feature. An enhancement of 2~dex results in an extremely strong and extended feature. The \ion{Ca}{II} H\&K feature is only weakly affected by the calcium within the shell at this epoch. The helium substrate requires more enhancement of calcium within the shell relative to that of the hydrogen (Figure \ref{fig:d2-H}) or carbon-oxygen (Figure \ref{fig:d2-CO}) substrates. A color figure is available online.  }
\end{figure}

Figure \ref{fig:d2-He} shows the spectra that result from shells with a helium substrate, again using the N100+0 abundance model for the ejecta. A purely solar abundance of calcium in the shell is insufficient to generate the observed CaNIR HVF due to the calcium being more likely to be in a higher ionization state (e.g. \ion{Ca}{III}) because of the higher ionization potential of helium and associated lower electron density in the gas. For He-Solar+0.0 model, the CaNIR feature is almost entirely photospheric, lacking evidence of a HVF. Calcium enhancement in the shell of 1~dex above solar levels produces a weak calcium HVF, but an enhancement of 2~dex above the solar value results in a very strong calcium feature. Note that the solar and +1 dex models show the feature around 7800\Ang{} that appeared in some of the hydrogen models (see Figure 
\ref{fig:d2-H}).

We have not attempted to generate a better fit for the shells with a helium substrate. We estimate that, for the N100+0 abundance model for the ejecta, a shell enhancement of about 1.3~dex will produce a reasonably good fit to the observed feature of SN~2011fe. Other than the CaNIR feature, the spectrum resulting from a shell with a helium substrate is identical to that of a shell with a hydrogen substrate, i.e. there are no helium absorption or emission features and the \ion{Ca}{II} H\&K feature is unaffected by calcium within the shell. 

\subsection{5 days}

\begin{figure}\centering
\includegraphics[height=\columnwidth,angle=-90]{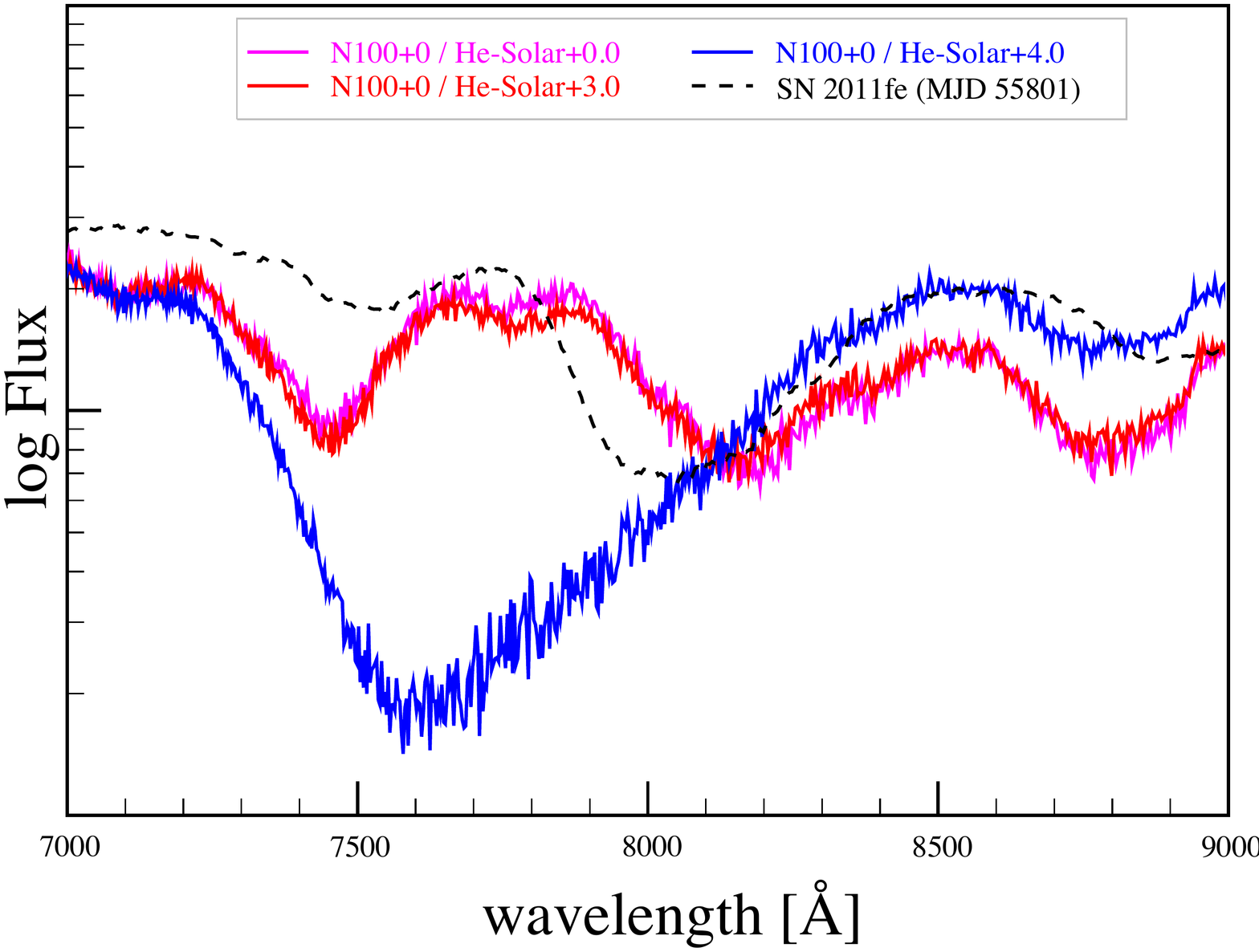}
\center
\includegraphics[height=\columnwidth,angle=-90]{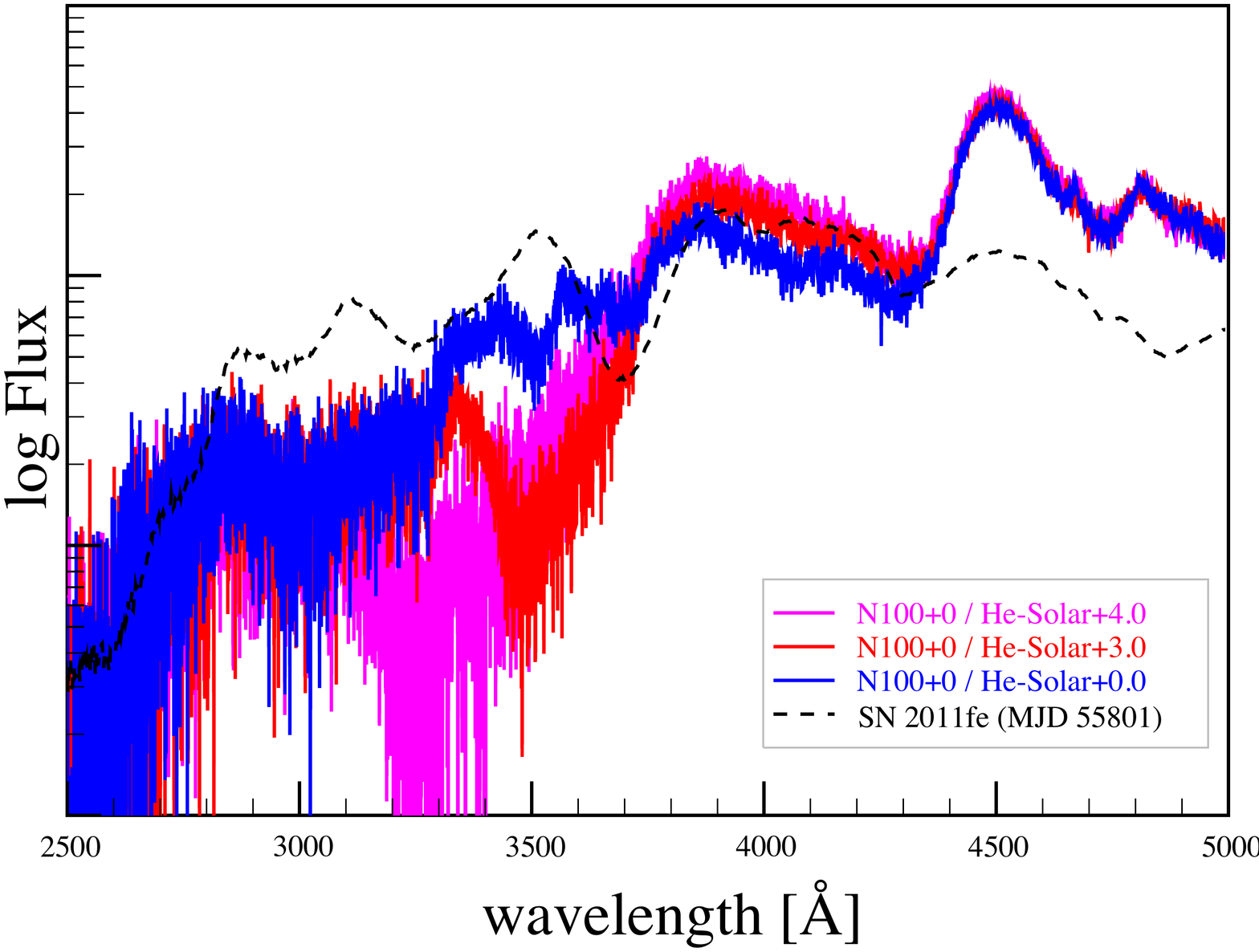}
\caption[He substrate shells at 5~d (CaNIR)]{\label{fig:d5-He} Top: Like Figure
\ref{fig:d5-H} at 5~d after the explosion, but for a shell with a helium substrate and calcium enhancement in the shell, relative to a solar value, ranging from 0 -- +4~dex, and focused on the CaNIR features. At 7800 \AA, the models from top to bottom represent N100+0/He-solar+0.0, N100+0/He-solar+3.0, and N100+0/He-solar+4.0, respectively. An abundance of +3~dex relative to solar is nearly indistinguishable from a purely solar abundance of helium at this epoch. Enhancement by an additional +1~dex results in an excessively strong feature. Bottom: Like 
the top panel but focused on the \ion{Ca}{II} H\&K feature. At 4000 \AA, the models from top to bottom represent N100+0/He-solar+4.0, N100+0/He-solar+3.0, and N100+0/He-solar+0.0, respectively. There is some enhancement of the feature already at 3~dex above a solar abundance of calcium within the shell.
A color figure is available online. }
\end{figure}

Figure \ref{fig:d5-He} shows the synthetic spectra at 5~d for the shells with a helium substrate. For calcium abundances in the shell at up to 3~dex above a solar value there is no effect upon the CaNIR feature, although there is a small effect upon the \ion{Ca}{II} H\&K feature, shown in the bottom panel of Figure \ref{fig:d5-He}. An enhancement of 4~dex above a solar value results in an extremely strong and extended CaNIR feature. 


\subsection{9 days}

\begin{figure}\centering
\includegraphics[height=\columnwidth,angle=-90]{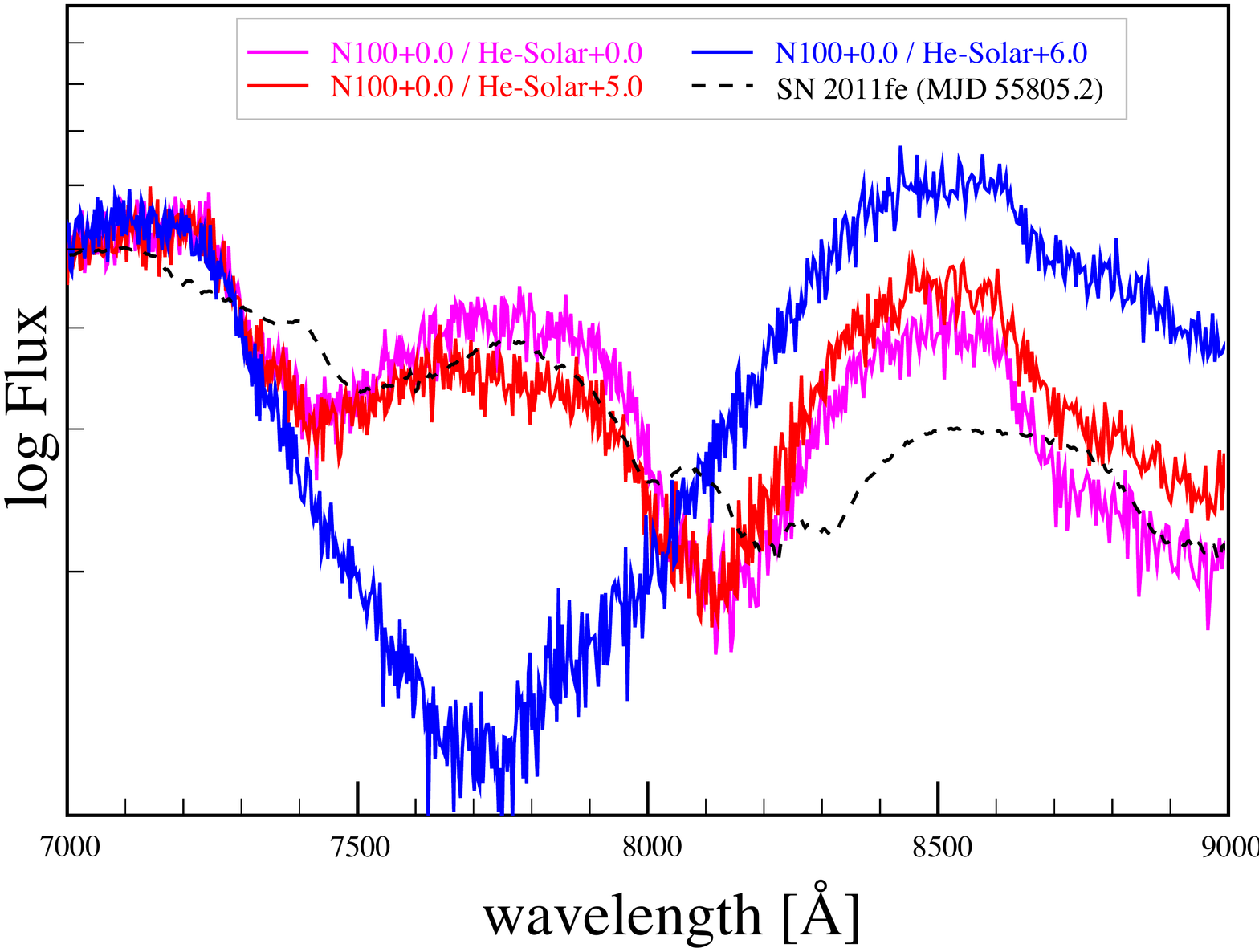}
\caption[He substrate shells at 9~d (CaNIR)]{\label{fig:d9-He} Like Figure 
\ref{fig:d9-H} at 9~d after the explosion but for a shell with a helium substrate with calcium enhancement ranging from 0 -- 6~dex in the shell. 
At 7800 \AA, the models from top to bottom represent N100+0/He-solar+0.0, N100+0/He-solar+5.0, and N100+0/He-solar+6.0, respectively. This order is reversed at 8500 \AA. The material within the shell has little to no effect on the spectrum at this epoch unless the calcium is enhanced by over 5~dex. Like the models with a hydrogen substrate, the PVF is slightly too blue. A color figure is available online.  }
\end{figure}

Figure \ref{fig:d9-He} shows the synthetic spectra at 9~d for the helium substrate abundance models. At this epoch, the material within the shell requires an enhancement of over 5~dex above solar value in order to have a significant effect upon the CaNIR feature. Like the models with a hydrogen substrate, enhancement of calcium within the shell tends to produce an excessive HVF, though an enhancement of over 5~dex is required for this to occur with the helium substrate. The CaNIR PVF tends to be slightly too blue. There is no evidence of helium absorption or emission features at this epoch.

\section{Solar-type abundance with carbon-oxygen substrate}\label{MZW1:apdx:CO}

\subsection{2 days}

\begin{figure}\centering
\includegraphics[height=\columnwidth,angle=-90]{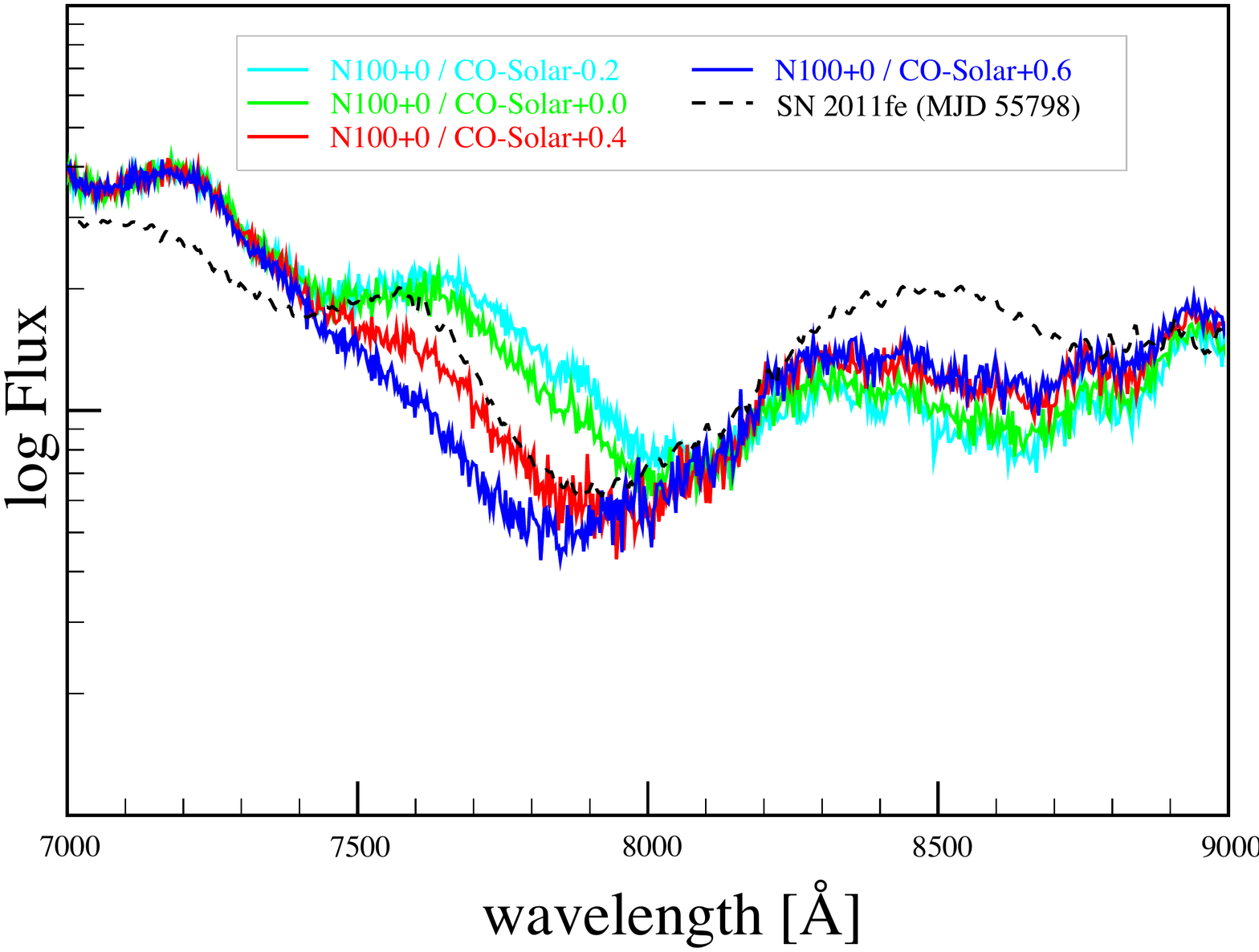}
\caption[CO substrate shells at 2~d (CaNIR)]{\label{fig:d2-CO} Like Figure \ref{fig:d2-H} at 2~d after the explosion, but for a shell with a carbon-oxygen substrate and calcium enhancement or depletion ranging from $-0.2$ -- +0.6~dex. At 7800 \AA, the models from top to bottom represent N100+0/CO-solar-0.2, N100+0/CO-solar+0.0, N100+0/CO-solar+0.4, and N100+0/CO-solar+0.6, respectively. These models tend to be very similar to those with a hydrogen substrate. An enhancement in calcium within the shell of +0.2 -- +0.4~dex provides a reasonably good fit to the observed feature in SN~2011fe. A color figure is available online.  }
\end{figure}

Figure \ref{fig:d2-CO} shows the spectra that result from shells with a carbon-oxygen substrate, again using the N100+0 abundance model for the ejecta. A purely solar abundance of calcium in the shell results in a feature that is similar to that observed in SN~2011fe. The models that include a carbon-oxygen substrate are nearly identical to those with a hydrogen substrate for the shell due to the similarity of ionization potential for hydrogen, carbon, and oxygen. Similar to the hydrogenic shell, a enhancement  of about $+$0.4~dex of calcium in the shell relative to a solar abundance will result in a reasonably good fit to the observed feature of SN~2011fe. There is no evidence of any effect upon carbon or oxygen absorption features due to the carbon and oxygen in the shell at this epoch.

\subsection{5 days}

\begin{figure}\centering
\includegraphics[height=\columnwidth,angle=-90]{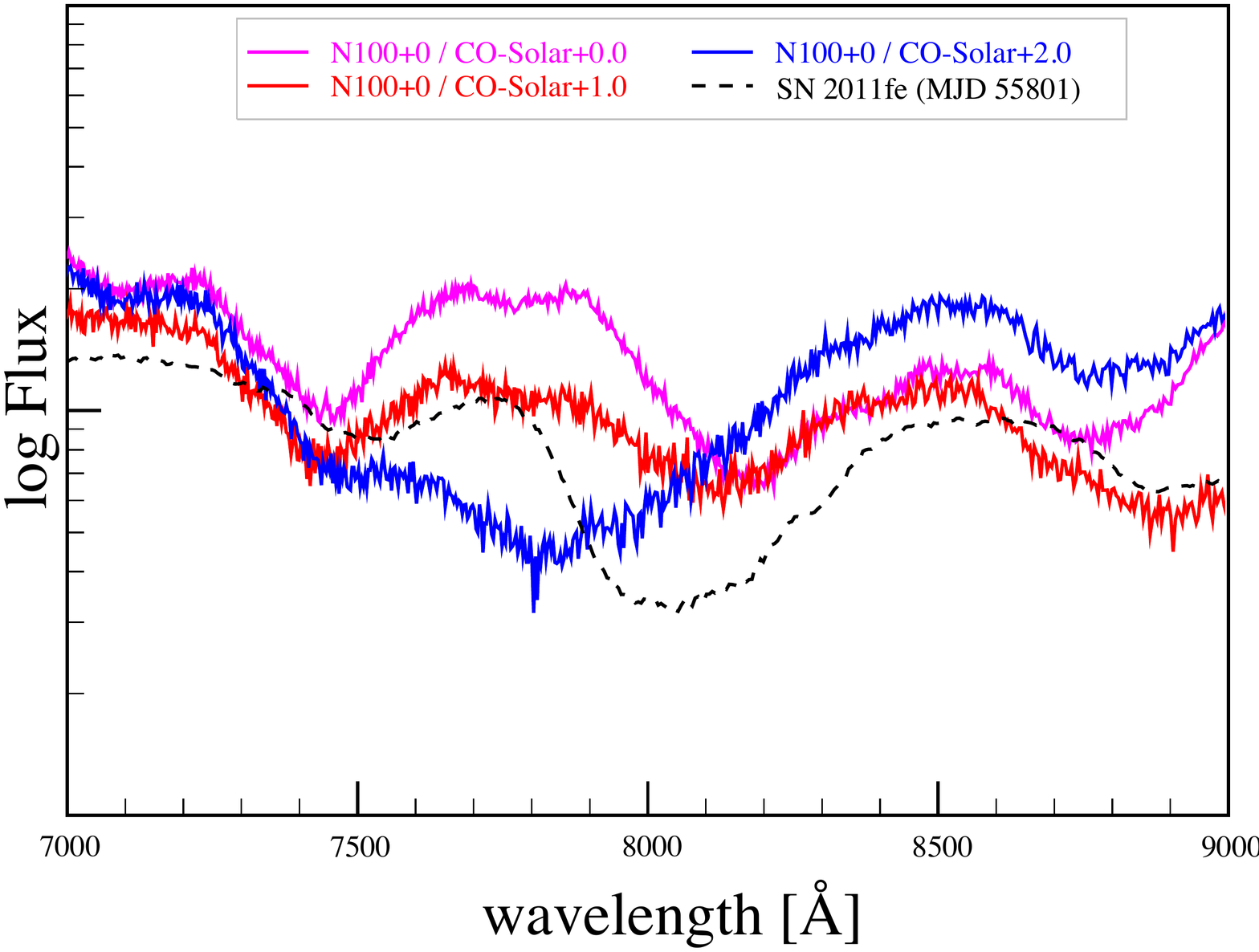}
\caption[CO substrate shells at 5~d (CaNIR)]{\label{fig:d5-CO} Like Figure \ref{fig:d5-H} at 5~d after the explosion, but for a shell with a carbon-oxygen substrate and calcium enhancement within the shell ranging from 0 -- +2~dex. At 7800 \AA, the models from top to bottom represent N100+0/CO-solar+0.0, N100+0/CO-solar+1.0, and N100+0/CO-solar+2.0, respectively.} At 0 -- +1~dex of enhancement of calcium within the shell, the HVF tend to be weak and extended in range. For this composition of the ejecta, an enhancement of +2~dex of calcium within the shell causes excessive absorption. A color figure is available online.
\end{figure}

Figure \ref{fig:d5-CO} shows the synthetic spectra at 5~d for a shell with a carbon-oxygen substrate with an enhancement in calcium within the shell of 0 -- 2~dex above solar value. For these models we see some evidence of a CaNIR HVF for all shells with solar abundance or more of calcium for the CaNIR feature. The CaNIR feature is clearly too strong for an abundance of 2~dex above a solar level of calcium. Like the models with a hydrogen substrate, the \ion{O}{I} 7773\Ang{} feature and \ion{Si}{II} 6355\Ang{} features are too strong due to the ejecta and are unaffected by material in the shell. 

\subsection{9 days}

\begin{figure}\centering
\includegraphics[height=\columnwidth,angle=-90]{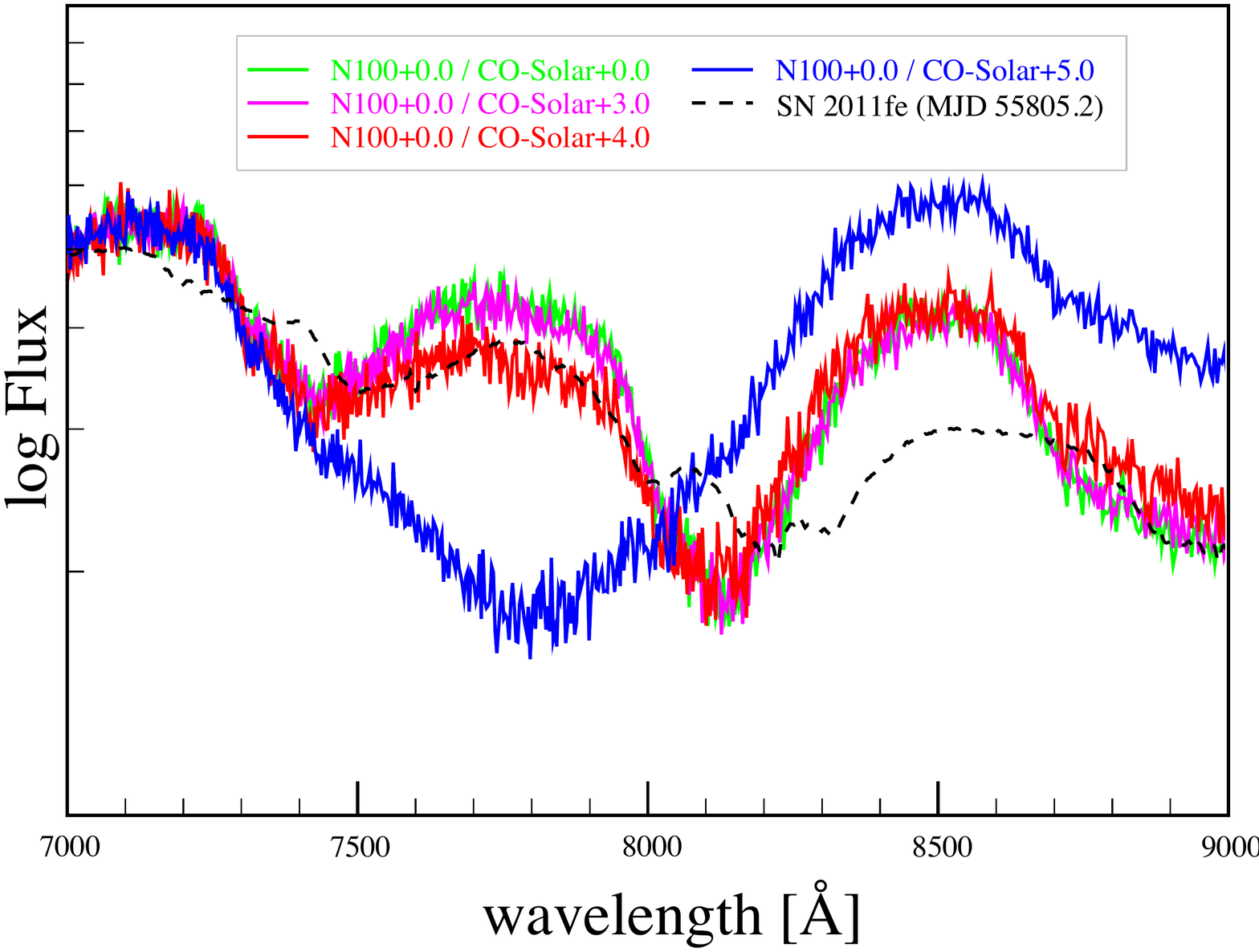}
\caption[CO substrate shells at 9~d (CaNIR)]{\label{fig:d9-CO} Like Figure \ref{fig:d9-H} at 9~d after the explosion, but for a shell with a carbon-oxygen substrate and calcium enhancement ranging from 0 -- 5~dex in the shell. 
At 7800 \AA, the models from top to bottom represent N100+0/CO-solar+0.0, N100+0/CO-solar+3.0 (very similar), N100+0/CO-solar+4.0, and N100+0/CO-solar+5.0, respectively.} At 3~dex of calcium enhancement or less, the material within the shell has little to no effect on the CaNIR feature at this epoch; there is slight evidence of a HVF with an enhancement of 4~dex, and a strong HVF at an enhancement of 5~dex. A color figure is available online.
\end{figure}

Figure \ref{fig:d9-CO} shows the synthetic spectra at 9~d for the models with a carbon-oxygen substrate, focused on the CaNIR feature. At this epoch, the feature is only weakly affected by the calcium within the shell, as long as the enhancement of the calcium is less than $\sim$4~dex. At 5~dex of enhancement, an excessively strong HVF is produced. At a purely solar abundance of calcium within the shell, this model is indistinguishable from that of a hydrogen or helium substrate with an equivalent amount of calcium.

\section*{Acknowledgments}
Thanks to Jeffrey Silverman, Jozsef Vink{\'o}, and G. H. Marion for many discussions of HVF and to Ken Shen for a discussion of his models. We thank Harriet Dinerstein for information and discussion of post-main sequence nucleosynthesis and the effect on calcium abundance. We are also grateful to the referee for a cogent and helpful report that served to improve the manuscript. Brian W. Mulligan was supported in part by the Graduate School Summer Fellowship at the University of Texas at Austin. J. Craig Wheeler was supported in part by the Samuel T. and Fern Yanagisawa Regents Professorship in Astronomy and by NSF grant 1813825. K.-C. Zhang was supported by the China Scholarship Council (CSC, NO. 201706210140). This research made use of \textsc{Tardis}, a community-developed software package for spectral synthesis in supernovae. The development of \textsc{Tardis} received support from the Google Summer of Code initiative and from ESA's Summer of Code in Space program. \textsc{Tardis} makes extensive use of Astropy and PyNE. For a current version of the code, see: https://zenodo.org/record/1292315.


\bibliographystyle{mnras}
\bibliography{MZW_2018_references}

\bsp	
\label{lastpage}
\end{document}